\documentclass[twocolumn,trackchanges]{aastex631}

\received{14 May 2023}
\revised{1 Oct 2023}
\accepted{3 Oct 2023}

\submitjournal{ApJ}

\shorttitle{Black Holes, Baryon Lifting, and Star-Formation Quenching}
\shortauthors{Voit et al.}

\begin{document}

\title{Black Hole Growth, Baryon Lifting, Star Formation, and IllustrisTNG}


\author[0000-0002-3514-0383]{G. Mark Voit}
\affiliation{Michigan State University,
Department of Physics and Astronomy,
East Lansing, MI 48824, USA}

\author[0000-0003-4754-6863]{Benjamin D. Oppenheimer}
\affiliation{University of Colorado, 
Center for Astrophysics and Space Astronomy, 389 UCB, Boulder, CO 80309, USA}

\author[0000-0002-5564-9873]{Eric F. Bell}
\affiliation{University of Michigan,
Department of Astronomy,
Ann Arbor, MI 48109, USA}

\author[0000-0001-5529-7305]{Bryan Terrazas}
\affiliation{Columbia University,
Department of Astronomy,
New York, NY 10027, USA}

\author[0000-0002-2808-0853]{Megan Donahue}
\affiliation{Michigan State University,
Department of Physics and Astronomy,
East Lansing, MI 48824, USA}




\begin{abstract}
Quenching of star formation in the central galaxies of cosmological halos is thought to result from energy released as gas accretes onto a supermassive black hole. The same energy source also appears to lower the central density and raise the cooling time of baryonic atmospheres in massive halos, thereby limiting both star formation and black hole growth, by lifting the baryons in those halos to greater altitudes. One predicted signature of that feedback mechanism is a nearly linear relationship between the central black hole's mass ($M_{\rm BH}$) and the original binding energy of the halo's baryons. We present the increasingly strong observational evidence supporting a such a relationship, showing that it extends up to halos of mass $M_{\rm halo} \sim 10^{14} M_\odot$. We then compare current observational constraints on the $M_{\rm BH}$--$M_{\rm halo}$ relation with numerical simulations, finding that black hole masses in IllustrisTNG appear to exceed those constraints at $M_{\rm halo} < 10^{13} M_\odot$ and that black hole masses in EAGLE fall short of observations at $M_{\rm halo} \sim 10^{14} M_\odot$. A closer look at IllustrisTNG shows that quenching of star formation and suppression of black hole growth do indeed coincide with black hole energy input that lifts the halo's baryons. However, IllustrisTNG does not reproduce the observed $M_{\rm BH}$--$M_{\rm halo}$ relation because its black holes gain mass primarily through accretion that does not contribute to baryon lifting. We suggest adjustments to some of the parameters in the IllustrisTNG feedback algorithm that may allow the resulting black hole masses to reflect the inherent links between black hole growth, baryon lifting, and star formation among the massive galaxies in those simulations.
\end{abstract}


\section{Introduction} 
\label{sec:intro}
 
A galaxy's star formation rate is tied to both the mass of its cosmological halo ($M_{\rm halo}$) and the mass of the black hole residing at its center ($M_{\rm BH}$). Large galaxy surveys spanning much of cosmic time show that the central galaxies of cosmological halos vigorously form stars until $M_{\rm halo}$ exceeds $\sim 10^{12} \, M_\odot$ \citep[e.g.,][]{Behroozi+2013ApJ...762L..31B,Behroozi2019MNRAS.488.3143B}. Star formation then subsides as $M_{\rm halo}$ increases toward $\sim 10^{13} \, M_\odot$. However, suppression of star formation among the central galaxies of present-day cosmological halos correlates more closely with the central velocity dispersion ($\sigma_v$) of a galaxy's stars than with $M_{\rm halo}$ \citep[e.g.,][]{Wake_2012ApJ...751L..44W,Bell_2012ApJ...753..167B,Woo_2015MNRAS.448..237W,Teimoorinia_2016MNRAS.457.2086T,Bluck_2016MNRAS.462.2559B,Bluck_2020MNRAS.492...96B}, implying that star-formation quiescence depends more directly on galactic structure than on halo mass. 

The mass of a galaxy's central black hole also closely correlates with $\sigma_v$ \citep[e.g.,][]{KormendyHo2013ARAA..51..511K}, suggesting a causal link between galaxy evolution and black hole growth. Observations show that suppression of star formation does indeed correlate with $M_{\rm BH}$ among nearby galaxies of similar stellar mass \citep{Terrazas_2016ApJ...830L..12T,Terrazas_2017ApJ...844..170T}. Both galactic structure and black hole growth therefore seem to conspire in the shutdown of star formation known as quenching.

Eruptions of feedback energy as a galaxy's central black hole grows are thought to be crucial for limiting black hole growth and perhaps also galactic star formation. An early analysis by \citet{SilkRees1998AA...331L...1S} proposed that the energy released as a galaxy's central black hole grows would limit the black hole's growth once it surpassed the energy required to lift baryons out of the galaxy's bulge, or perhaps even out of the galaxy's entire potential well \citep[see also][]{Haehnelt_1998MNRAS.300..817H}. The predicted result: a scaling relation ($M_{\rm BH} \propto \sigma_v^5$) similar to the observed one \citep[for a more recent review of similar ideas, see][]{KingPounds2015ARA&A..53..115K}. 

A causal connection between $M_{\rm BH}$ and $M_{\rm halo}$ became more plausible when \citet{Ferrarese_2002ApJ...578...90F} showed that nearby spiral galaxies follow the same $M_{\rm BH}$--$M_{\rm halo}$ scaling relation as their more massive elliptical counterparts, based on assuming that a galaxy's central velocity dispersion and rotation speed are proportional to each other. When $M_{\rm halo}$ is defined in terms of a mean matter density, the halo's circular velocity is $v_{\rm c} \propto M_{\rm halo}^{1/3}$, the specific binding energy of its matter is $E_{\rm B} / M_{\rm halo} \propto v_{\rm c}^2$, and the total binding energy is $E_{\rm B} \propto M_{\rm halo} v_{\rm c}^2 \propto v_{\rm c}^5$. Black-hole growth limited by baryon lifting should therefore result in $M_{\rm BH} \propto \sigma_v^5 \propto M_{\rm halo}^{5/3}$. The \citet{Ferrarese_2002ApJ...578...90F} data set supported the baryon lifting hypothesis because it indicated $M_{\rm BH} \propto M_{\rm halo}^\gamma$ with $\gamma \approx 1.65$--1.82, depending on the methods used to infer $M_{\rm halo}$ from $\sigma_v$ and $v_{\rm c}$. A few years later, \cite{Bandara_2009ApJ...704.1135B} strengthened the evidence for such a power-law relation, through a survey that used lensing observations to obtain $M_{\rm halo}$ and indirectly inferred $M_{\rm BH}$ from $\sigma_v$, finding $M_{\rm BH} \propto M_{\rm halo}^{1.55 \pm 0.31}$. However, some experts remained deeply skeptical of a direct causal connection between $M_{\rm BH}$ and $M_{\rm halo}$ \citep[e.g.,][]{KormendyBender_2011Natur.469..377K,KormendyHo2013ARAA..51..511K}.

Cosmological simulations then put the proposed relationship between $M_{\rm BH}$ and $M_{\rm halo}$ on a firmer theoretical footing. \citet{BoothSchaye_2010MNRAS.405L...1B} demonstrated that energy released by the black-hole feedback algorithm in their simulations led to the scaling relation $M_{\rm BH} \propto M_{\rm halo}^{1.55 \pm 0.05}$, with a normalization coefficient proportional to the assumed ratio of accreted mass to energy output. This relationship arose because the simulated black holes grew through accretion until they released an energy comparable to the gravitational binding energy of \textit{all} the halo's baryons. The accumulating energy then lifted the halo's baryons, thereby lowering the density, pressure, and cooling time of baryons in the black hole's vicinity, reducing its long-term accretion rate and limiting its growth. The resulting power-law slope ended up slightly smaller than the 5/3 prediction for identically structured halos because the dark-matter density profiles of lower-mass halos tend to be more centrally concentrated than those of higher-mass halos, leading to a shallower dependence of specific binding energy on halo mass.

More recent simulations incorporating many more astrophysical details have demonstrated that lifting of a halo's baryons via black-hole feedback may also be critical for suppressing star formation \citep{Davies_2019MNRAS.485.3783D,Davies_2020MNRAS.491.4462D,
Oppenheimer_2020MNRAS.491.2939O,Terrazas2020MNRAS.493.1888T,Zinger_2020MNRAS.499..768Z,Appleby_2021MNRAS.507.2383A}. In simulated halos with $M_{\rm halo} \gtrsim 10^{12} M_\odot$, black-hole feedback is the prime mover of baryons beyond the virial radius. Furthermore, simulated galaxies centered within halos of mass $\sim 10^{12} M_\odot$ tend to have star formation rates that correlate with the proportion of the halo's baryons remaining within the virial radius.

A similar story has emerged from analyses of correlations between star-formation quenching, the structural properties of galaxies, and the masses of their central black holes. According to \citet{ChenFaber_2020ApJ...897..102C}, the $M_{\rm BH}$--$\sigma_v$ relation among galaxies with active star formation has a power-law slope similar to the $M_{\rm BH}$--$\sigma_v$ relation among quiescent galaxies but a mass normalization approximately an order of magnitude smaller at fixed $\sigma_v$. The transition from active to quenched star formation therefore appears to be associated with rapid black hole mass growth. It is also consistent with an amount of black hole growth that is proportional to the halo's baryonic binding energy, suggesting that quenching results from lifting of the halo's baryons via black hole feedback.\footnote{
In the interpretation presented by \citet{ChenFaber_2020ApJ...897..102C}, the amount of injected feedback energy needed to quench star formation is 4 times the halo's baryonic binding energy, but the numerical factor is degenerate with the conversion efficiency of accreted mass-energy to feedback energy, which they assume to be $0.01$.}

The proposed connection between baryon lifting and quenching of star formation is theoretically appealing, but then why does quiescence correlate more closely with $\sigma_v$ than with $M_{\rm halo}$? \citet{Voit_2020ApJ...899...70V} have argued that baryon lifting via black hole feedback is an inevitable consequence of structural evolution that raises a galaxy's central stellar mass density, as reflected by $\sigma_v$. The central cooling rate of hot gas in galaxies with large $\sigma_v$ depends primarily on circumgalactic gas pressure. Consequently, as $\sigma_v$ rises above a critical value determined by stellar heating, black hole fueling becomes linked to circumgalactic pressure. Once that link is established, $M_{\rm BH}$ then grows to depend directly on $M_{\rm halo}$ as cumulative black hole energy injection rises to scale with the halo's baryonic binding energy \citep[for an extensive review, see][]{DonahueVoit2022PhR...973....1D}.

\begin{figure*}[th]
\begin{center}
\includegraphics[width=6.9in,trim=0.2in -0.1in 0.0in 0.0in]{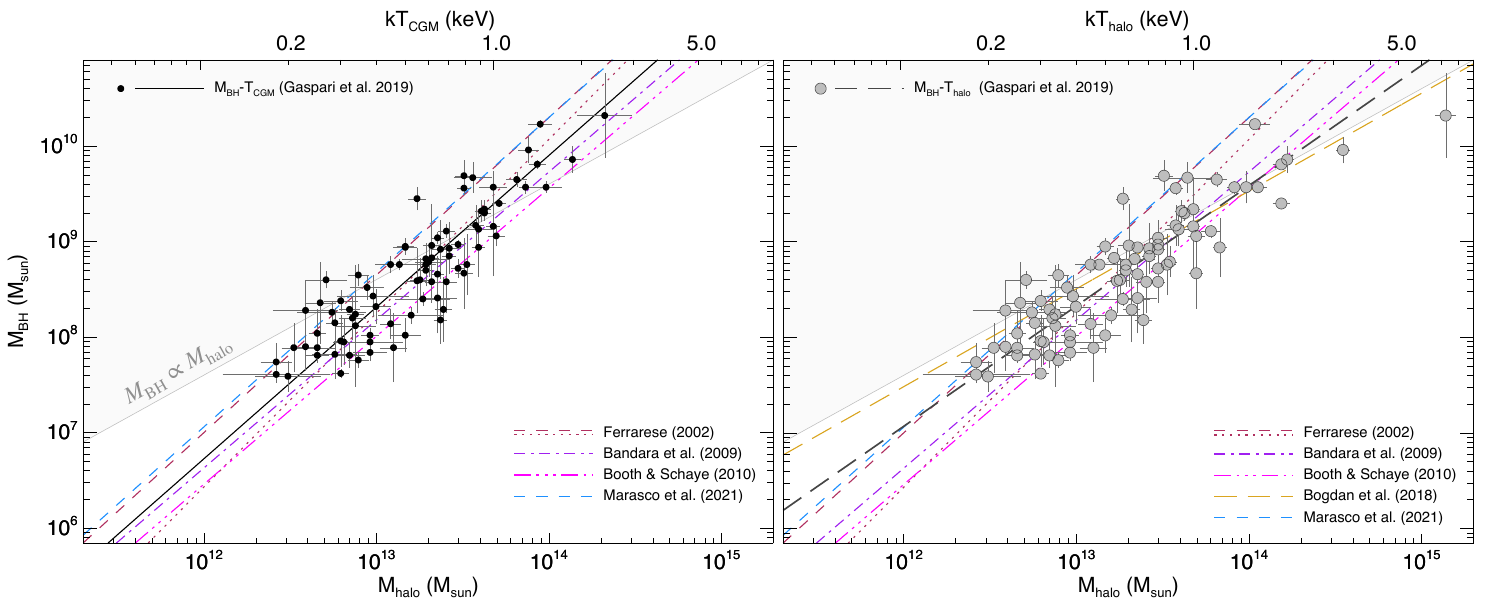}
\end{center}
\caption{Observed relationships between black hole mass ($M_{\rm BH}$), halo mass ($M_{\rm halo}$), and atmospheric temperature ($T$). Dotted and dashed red lines show two $M_{\rm BH}$--$M_{\rm halo}$ relations from \citet{Ferrarese_2002ApJ...578...90F}. Dot-dashed purple lines show the $M_{\rm BH}$--$M_{\rm halo}$ relation from \citet{Bandara_2009ApJ...704.1135B}. Dot-dot-dot-dashed magenta lines show the simulated $M_{\rm BH}$--$M_{\rm halo}$ relation from \citet{BoothSchaye_2010MNRAS.405L...1B}. Dashed blue lines show the $M_{\rm BH}$--$M_{\rm halo}$ relation from \citet{Marasco_2021MNRAS.507.4274M}. Solid grey lines separating shaded from unshaded regions indicate a linear $M_{\rm BH}$--$M_{\rm halo}$ correlation.
Black points in the left panel show the $M_{\rm BH}$--$T_{\rm CGM}$ measurements from \citet{Gaspari2019}. The best fitting power law relation ($M_{\rm BH} \propto M_{\rm halo}^{1.6}$) shown by the solid black line is clearly super-linear. Grey points in the right panel show $M_{\rm BH}$--$T_{\rm halo}$ measurements from \citet{Gaspari2019}, and a dashed black line shows the best fitting power law relation ($M_{\rm BH} \propto M_{\rm halo}^{1.3}$).  An additional dashed gold line in the right panel shows the $M_{\rm BH}$--$T_{\rm halo}$ relation from the sample of \citet{Bogdan_2018ApJ...852..131B}. In both panels, the relationship  $kT_{\rm X} = 6 \, {\rm keV} \times (M_{\rm 200c} / 10^{15} \, M_\odot)^{1.7}$ maps gas temperature onto halo mass. However, masses based on $T_{\rm CGM}$ (left panel) are underestimates in cases where $T_{\rm halo} \gg T_{\rm CGM}$.
\vspace*{2em}
\label{fig:MBH-TCGM-Thalo}}
\end{figure*}

This paper presents evidence favoring such a three-way link between black hole growth, baryon lifting, and star-formation quiescence. Section \ref{sec:MBH-Mhalo} starts things off by examining current observational assessments of the $M_{\rm BH}$--$M_{\rm halo}$ relation, comparing them with the results of numerical simulations, and finding general support for the three-way link, except in IllustrisTNG, which requires a deeper examination. Section~\ref{sec:lifting} establishes that quenching of star formation in IllustrisTNG does, in fact, coincide with kinetic feedback input sufficient to lift a halo's baryons. Section~\ref{sec:BHmode} analyzes the contrasting roles that the thermal (``quasar") and kinetic (``radio") feedback modes of IllustrisTNG play in baryon lifting. Section~\ref{sec:price} briefly discusses how the feedback efficiency parameters employed in numerical simulations determine the ``price" of feedback, as reflected by black hole mass growth. Section~\ref{sec:priceisright} speculates about how ``price" changes might bring IllustrisTNG black hole masses into better agreement with both observations and the predicted $M_{\rm BH}$--$M_{\rm halo}$ scaling relation. Section~\ref{sec:summary} summarizes the paper's findings. 



\section{Black Holes and Halo Masses}
\label{sec:MBH-Mhalo}

The introduction mentioned some of the observational constraints on the $M_{\rm BH}$--$M_{\rm halo}$ relation. Now we will take a closer look at those observations and compare them with what emerges from the IllustrisTNG and EAGLE cosmological simulations. We will focus most closely on the halo mass range from $10^{12.5} M_\odot$ to $10^{14} M_\odot$, because that is where X-ray observations provide both direct evidence for baryon lifting and reliable estimates of $M_{\rm halo}$. Our review of the literature is therefore neither comprehensive nor complete.

\newpage

\subsection{Observations}

Figure \ref{fig:MBH-TCGM-Thalo} illustrates several relationships between $M_{\rm BH}$ and $M_{\rm halo}$. A dotted red line shows the relation
\begin{equation}
    M_{\rm BH} = 10^{8.25} M_\odot 
        \left( \frac {M_{\rm halo}} {10^{13} M_\odot} \right)^{1.82}
\end{equation}
corresponding to equation (4) from \citet{Ferrarese_2002ApJ...578...90F}. A dashed red line shows the relation
\begin{equation}
    M_{\rm BH} = 10^{8.50} M_\odot 
        \left( \frac {M_{\rm halo}} {10^{13} M_\odot} \right)^{1.65}
\end{equation}
corresponding to equation (6) from \citet{Ferrarese_2002ApJ...578...90F}. A dot-dashed purple line shows the relation 
\begin{equation}
    M_{\rm BH} = 10^{8.18} M_\odot 
        \left( \frac {M_{\rm halo}} {10^{13} M_\odot} \right)^{1.55}
\end{equation}
from \citet{Bandara_2009ApJ...704.1135B}. And a dashed blue line shows the relation \begin{equation}
    M_{\rm BH} = 10^{8.68} M_\odot 
        \left( \frac {M_{\rm halo}} {10^{13} M_\odot} \right)^{1.62}
\end{equation}
derived from observations compiled by \cite{Marasco_2021MNRAS.507.4274M}.

Those four assessments of the $M_{\rm BH}$--$M_{\rm halo}$ relation generally align with each other and also with the dot-dot-dot-dashed magenta line showing the relation 
\begin{equation}
    M_{\rm BH} = 10^{8.01} M_\odot 
        \left( \frac {M_{\rm halo}} {10^{13} M_\odot} \right)^{1.55}
\end{equation}
that \cite{BoothSchaye_2010MNRAS.405L...1B} found in cosmological numerical simulations of black hole feedback. However, \citet{Ferrarese_2002ApJ...578...90F} inferred $M_{\rm halo}$ from galactic dynamics, not halo properties. \citet{Bandara_2009ApJ...704.1135B} inferred $M_{\rm BH}$ from $\sigma_v$, not direct dynamical measurements of $M_{\rm BH}$. And \citet{Marasco_2021MNRAS.507.4274M} used a heterogeneous set of proxies for $M_{\rm halo}$, making it difficult to assess the impact of systematic uncertainties on their best fitting $M_{\rm BH}$ normalization.

X-ray analyses have recently provided more direct constraints on the $M_{\rm BH}$--$M_{\rm halo}$ relation \citep{Bogdan_2018ApJ...852..131B,Lakhchuara_2019MNRAS.488L.134L,
Gaspari2019}. Among those analyses, the \citet{Gaspari2019} sample relies on the largest data set (85 galaxies with dynamical measurements of $M_{\rm BH}$). Where possible, that data set provides two distinct X-ray temperatures, one ($T_{\rm CGM}$) measured within a few effective radii of the galaxy and another ($T_{\rm halo}$) more representative of the halo gas at larger radii \citep[for details, see][]{Gaspari2019}.\footnote{In our notation, $T_{\rm CGM}$ corresponds to their $T_{\rm x,g}$ and $T_{\rm halo}$ corresponds to their $T_{\rm x,c}$.} 

\begin{figure*}[th]
\begin{center}
\includegraphics[width=5.2in,trim=0.2in 0.0in 0.0in 0.0in]{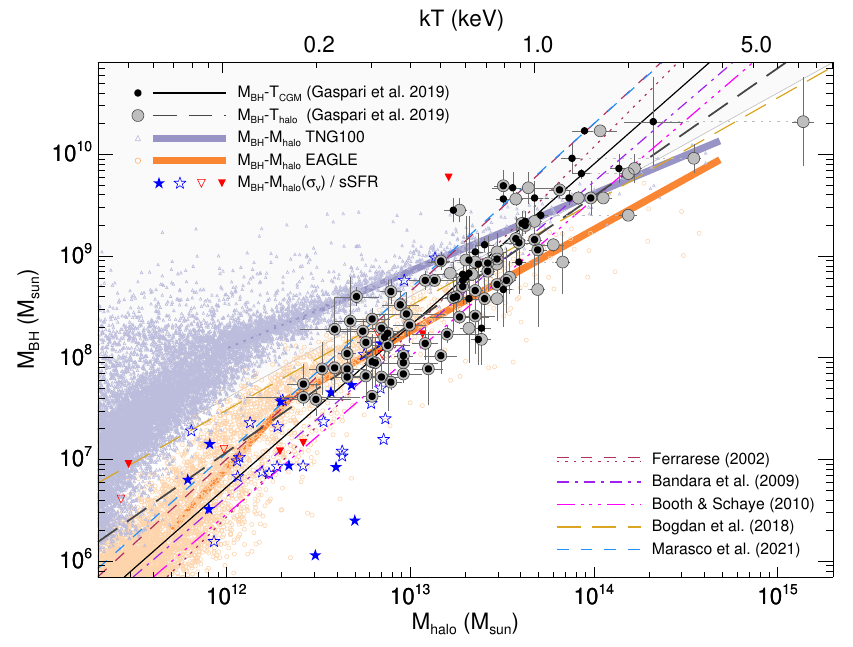}
\end{center}
\caption{Comparisons between numerical simulations and the $M_{\rm BH}$--$M_{\rm halo}$ relations inferred from observations. All lines and symbols in common with Figure~\ref{fig:MBH-TCGM-Thalo} represent identical quantities. Black and grey points representing the same black hole from \citet{Gaspari2019} are connected by horizontal dotted lines. Blue stars and inverted red triangles represent galaxies from \citet{Terrazas_2017ApJ...844..170T}, with shapes and shading encoding sSFR as described in \S \ref{sec:MBH-Mhalo}, and rely on the $M_{\rm halo} (\sigma_v)$ relation in equation (\ref{eq:Mhalo-sigmav}). Purple triangles show $M_{\rm BH}$ and $M_{\rm halo}$ from the TNG100 simulation, with a thick purple line illustrating the power-law fit for $M_{\rm halo} > 10^{12} M_\odot$ from \citet{Truong_2021MNRAS.501.2210T}. Small orange circles show $M_{\rm BH}$ and $M_{\rm halo}$ from the EAGLE simulation, with a thick orange line illustrating a piecewise power-law fit to EAGLE. 
\vspace*{2em}
\label{fig:MBH_Mhalo}}
\end{figure*}

Interestingly, \citet{Gaspari2019} found that $M_{\rm BH}$ correlates more closely with $T_{\rm CGM}$ than with any other observable property, including even $\sigma_v$, among a large set of observable galactic and X-ray characteristics. Black circles in the left panel of Figure~\ref{fig:MBH-TCGM-Thalo} show that correlation, with X-ray temperature mapped onto $M_{\rm halo}$ using a relationship
\begin{equation}
    M_{\rm 200 c} = 10^{15} M_\odot  \, \left( \frac {kT_{\rm X}} {6 \, {\rm keV}} \right)^{1.7}
\end{equation}
based on observations by \citet{Sun+09}. The original $M_{\rm halo}$--$T_{\rm X}$ relation used X-ray data to derive the mass $M_{\rm 500c}$ within a radius encompassing a mean mass density 500 times the universe's critical density. Here, we have recalibrated it by setting $M_{\rm 200c} = 1.5 M_{\rm 500c}$, where $M_{\rm 200c}$ is defined using a density contrast of 200 instead of 500.\footnote{This conversion factor assumes a Navarro-Frenk-White mass profile with a concentration parameter $c_{200} \approx 4$ \citep[e.g.,][]{Merten_2015ApJ...806....4M}.} The resulting $M_{\rm BH}$--$M_{\rm halo}$ relation
\begin{equation}
    M_{\rm BH} = 10^{8.3} M_\odot 
        \left( \frac {M_{\rm halo}} {10^{13} M_\odot} \right)^{1.6}
\end{equation}
aligns well with the earlier but less direct constraints (see Figures~\ref{fig:MBH-TCGM-Thalo} and \ref{fig:MBH_Mhalo}). 

Applying the same $M_{\rm 200c}$--$T_{\rm X}$ relation to $T_{\rm halo}$ leads to a set of points (grey circles in the right panel of Figure \ref{fig:MBH-TCGM-Thalo}) that significantly depart from the $M_{\rm 200c}$--$T_{\rm CGM}$ relation above $\sim 1.5 \, {\rm keV}$, where $M_{\rm halo} \gtrsim 10^{14} M_\odot$ and $M_{\rm BH} \gtrsim 5 \times 10^9 M_\odot$. Apparently, the power-law slope of the $M_{\rm BH}$--$M_{\rm halo}$ relation flattens as halos go from the group scale to the cluster scale. The $M_{\rm BH}$--$T_{\rm halo}$ relation from \citet{Bogdan_2018ApJ...852..131B} supports this conclusion. Their sample spans a narrower mass range than the \citet{Gaspari2019} sample, is dominated by high-mass halos, and obtains an $M_{\rm BH}$--$T_{\rm halo}$ relation (dashed gold line in the right panel of Figure \ref{fig:MBH-TCGM-Thalo}) that is less steep than the \citet{Gaspari2019} $M_{\rm BH}$--$T_{\rm halo}$ relation (dashed black line in the right panel of Figure \ref{fig:MBH-TCGM-Thalo}).

The apparent leveling of the $M_{\rm BH}$--$M_{\rm halo}$ relation above $M_{\rm halo} \sim 10^{14} M_\odot$ may result from a qualitative change in how supermassive black holes interact with their environments, for two reasons: (1) it coincides with the mass scale at which halos appear to retain nearly all of their baryons, and (2) it coincides with the mass scale at which the circular velocity of a central galaxy no longer reflects the circular velocity of its dark-matter halo. Observational inventories of baryons in galaxy groups ($10^{13} M_\odot \lesssim M_{\rm halo} \lesssim 10^{14} M_\odot$) show that they contain only about half the cosmic baryon fraction \citep[e.g.,][]{Sun+09,Lovisari_2015A&A...573A.118L,Eckert_2021Univ....7..142E}, while similar inventories of galaxy clusters ($M_{\rm halo} \gtrsim 10^{14.5} M_\odot$) find essentially all of the expected baryons \citep[e.g.,][]{Pratt_2009_REXCESS_LX-T}. Among galaxy clusters, radiative losses plausibly balance the black hole's energy input \citep[e.g.,][]{mn07,McNamaraNulsen2012NJPh...14e5023M}, but the baryon lifting observed in lower mass halos implies that black hole power, when integrated over time, greatly exceeds cumulative radiative losses \citep{DonahueVoit2022PhR...973....1D}. Leveling of the $M_{\rm BH}$--$M_{\rm halo}$ relation therefore appears to happen where black hole power is no longer capable of lifting a halo's baryons and instead dissipates through radiative losses. Furthermore, the pronounced differences between $T_{\rm CGM}$ and $T_{\rm halo}$ observed among galaxy clusters reflect a disruption of the usual link between the circular velocity of a cosmological halo and the circular velocity of its central galaxy. In galaxy groups, $T_{\rm CGM}$ and $T_{\rm halo}$ are typically more similar because the circular velocity of a group's potential well is closer to the circular velocity of its central galaxy.

\citet{DonahueVoit2022PhR...973....1D} have hypothesized that $M_{\rm BH}$ is more highly correlated with $T_{\rm CGM}$ than with $T_{\rm halo}$ because it more closely represents the halo's baryonic binding energy \textit{at the time black hole feedback lifted those baryons and quenched the central galaxy's star formation}. For example, the most massive black hole in the \citet{Gaspari2019} sample resides in NGC~4889, the central galaxy of the Coma Cluster. Its atmospheric temperature ($T_{\rm CGM} \approx 2.4 \, {\rm keV}$) is considerably lower than the cluster's atmospheric temperature ($T_{\rm halo} \approx 7.2 \, {\rm keV}$), which indicates $M_{\rm halo} \sim 10^{15} \, M_\odot$. Presumably, the baryon-lifting event that quenched star formation in NGC~4889 predated the Coma Cluster's growth to such a large mass, explaining why its black hole's mass falls below the power-law $M_{\rm BH}$--$M_{\rm halo}$ relation followed by lower-mass halos in the right panel of Figure~\ref{fig:MBH-TCGM-Thalo}.

X-ray assessments of the $M_{\rm BH}$--$M_{\rm halo}$ relation become increasingly difficult as $M_{\rm halo}$ drops below $\sim 10^{13} M_\odot$ because there are fewer and fewer X-ray photons for making temperature measurements. Also, the link between $M_{\rm halo}$ and X-ray temperatures measurements may become weaker because of transient temperature fluctuations produced by feedback events \citep[e.g.,][]{Truong_2021MNRAS.501.2210T}. However, the $M_{\rm BH}$--$M_{\rm halo}$ relation can be extended toward lower masses using other mass proxies. 

As an example, the blue stars and inverted red triangles in Figure \ref{fig:MBH_Mhalo} show an extension based on $\sigma_v$ assuming
\begin{equation}
    M_{\rm halo} = 10^{12.9} M_\odot 
        \left( \frac {\sigma_v} {200 \, {\rm km \, s^{-1}}} \right)^3
        \label{eq:Mhalo-sigmav}
\end{equation}
which is equivalent to $M_{\rm 200c}$ for a singular isothermal sphere with an isotropic velocity dispersion identical to the galaxy's observed $\sigma_v$. The stars and triangles represent galaxies from the \citet{Terrazas_2017ApJ...844..170T} sample that do not appear in the \citet{Gaspari2019} sample. Their shapes and shading represent specific star-formation rates (sSFR) equal to each galaxy's star-formation rate ($\dot{M}_*$) divided by its stellar mass ($M_*$):
\begin{itemize}
    \item Filled stars, 
                $\geq 10^{-10.3} \, {\rm yr}^{-1}$
    \item Open stars, 
                $10^{-11.0} {\rm yr}^{-1}$ to $10^{-10.3} \, {\rm yr}^{-1}$
    \item Open triangles, 
                $10^{-11.7} {\rm yr}^{-1}$ to $10^{-11} \, {\rm yr}^{-1}$
    \item Filled triangles, 
                $\leq 10^{-11.7} \, {\rm yr}^{-1}$ .
\end{itemize}
We will return to the significance of sSFR in \S \ref{sec:Proxies-Mstar}. For now, we will simply note that those points align with the $M_{\rm BH}$--$T_{\rm CGM}$ relation.

\subsection{Simulations}

Figure \ref{fig:MBH_Mhalo} shows how the EAGLE \citep{Schaye_EAGLE_2015MNRAS.446..521S} and TNG100 \citep{Pillepich_EAGLE_2018MNRAS.475..648P,Nelson_2018_bimodality} simulations compare with the observations. The EAGLE points (orange circles) overlap with the observational points up to $M_{\rm 200c} \sim 10^{13.5} M_\odot$ but predict smaller black hole masses in more massive halos. A power-law fit to the EAGLE points with $M_{\rm 200c} > 10^{12.3} M_\odot$ gives
\begin{equation}
        M_{\rm BH} = 10^{8.1} M_\odot 
        \left( \frac {M_{\rm 200c}} {10^{13} M_\odot} \right)^{1.0}
        \; \; ,
\end{equation}
but the EAGLE power-law slope is steeper at lower halo masses \citep{Rosas-Guevara_2016MNRAS.462..190R}. For example, the best fitting power law for $10^{11.5} M_\odot < M_{\rm 200c} < 10^{12.3} M_\odot$ is $M_{\rm BH} \propto M_{\rm halo}^{2.0}$. A thick orange line in Figure \ref{fig:MBH_Mhalo} illustrates the two pieces of this piecewise power-law fit. The high-mass flattening of the EAGLE relation qualitatively agrees with observations but sets in nearer to $M_{\rm halo} \sim 10^{12.3} M_\odot$ than to $\sim 10^{14} M_\odot$. Interestingly, fitting all of the EAGLE points having $M_{\rm 200c} > 10^{11.5} M_\odot$ with a single power law yields a relation with essentially the same slope found by \citet{BoothSchaye_2010MNRAS.405L...1B} but a slightly greater $M_{\rm BH}$ normalization.

The IllustrisTNG points (small purple triangles) representing $M_{\rm BH}$--$M_{\rm 200c}$ are less well aligned with the observational constraints. A thick purple line shows the power-law fit 
\begin{equation}
        M_{\rm BH} = 10^{8.8} M_\odot 
        \left( \frac {M_{\rm 200c}} {10^{13} M_\odot} \right)^{0.76}
\end{equation}
from \citet{Truong_2021MNRAS.501.2210T} for halos with $M_{\rm 200c} > 10^{12} M_\odot$. It agrees with the $M_{\rm BH}$ observations among the most massive halos ($M_{\rm halo} \gtrsim 10^{14} M_\odot$) but exceeds them at $M_{\rm halo} \lesssim 10^{13.5} M_\odot$, ending up near $M_{\rm BH} \sim 10^8 M_\odot$ at $M_{\rm halo} \sim 10^{12} M_\odot$. The anomalously large IllustrisTNG black hole masses at $M_{\rm halo} \sim 10^{12} M_\odot$ have previously been noted by \citet{Li_2020ApJ...895..102L}, in the context of the $M_{\rm BH}$--$\sigma_v$ relation, and by both \citet{Terrazas_2016ApJ...830L..12T,Terrazas_2017ApJ...844..170T} and \citet{Habouzit_2021MNRAS.503.1940H}, in the context of the $M_{\rm BH}$--$M_*$ relation.

\subsection{Accretion versus Mergers}

Black holes in IllustrisTNG halos above $M_{\rm halo} \sim 10^{12} M_\odot$ accumulate mass primarily through mergers with other black holes \citep{Weinberger_2018MNRAS.479.4056W}. Merger-dominated growth therefore results in a sub-linear $M_{\rm BH}$--$M_{\rm halo}$ relation \citep{Truong_2021MNRAS.501.2210T}. However, Figure \ref{fig:MBH-TCGM-Thalo} shows that the observed $M_{\rm BH}$--$T_{\rm CGM}$ relation indicates that the $M_{\rm BH}$--$M_{\rm halo}$ relation is super-linear up to $M_{\rm halo} \sim 10^{14} M_\odot$. Those observations therefore imply either (1) that $M_{\rm BH}$ grows in proportion to $M_{\rm halo}^{1.6}$ as halo mass evolves up to $\sim 10^{14} M_\odot$, or (2) that black holes in halos that will eventually merge to form a $\sim 10^{14} M_\odot$ halo grow through accretion to greater masses than black holes forming in halos destined to reach lower halo masses. This latter possibility would imply that black-hole accretion early in time is influenced by environmental effects extending beyond the borders of its own cosmological halo. The implications of EAGLE's nearly linear $M_{\rm BH}$--$M_{\rm halo}$ relation at high masses are less clear and may indicate a combination of merger-driven and accretion-driven growth beyond $M_{\rm halo} \sim 10^{12.3} M_\odot$. 

\subsection{Halo Mass Proxies}
\label{sec:Proxies}

\begin{figure*}[th]
\begin{center}
\includegraphics[width=6.8in,trim=0.2in 0.0in 0.0in 0.0in]{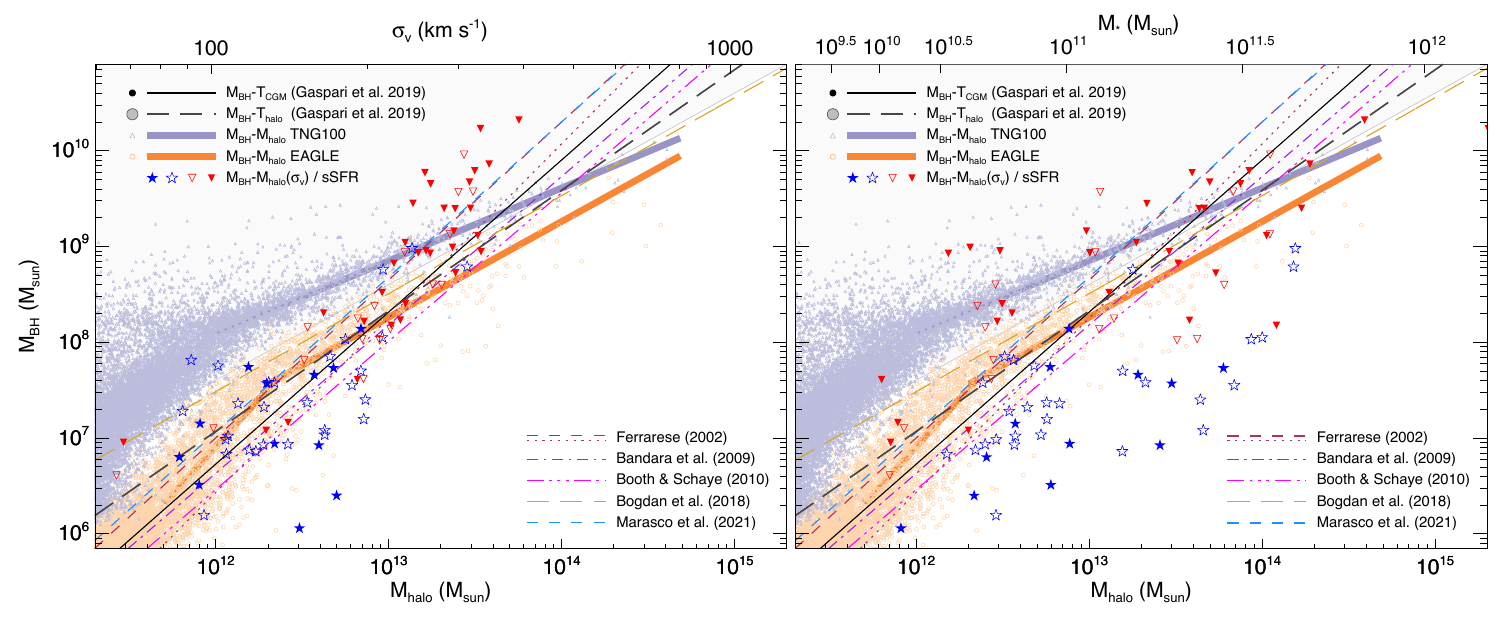}
\end{center}
\caption{Relationships between $M_{\rm BH}$ and $M_{\rm halo}$ inferred from mass proxies other than X-ray temperature. Left panel: The $M_{\rm BH}$--$M_{\rm halo}$ relation inferred from $M_{\rm halo} (\sigma_v)$ using equation (\ref{eq:Mhalo-sigmav}). Right panel: The $M_{\rm BH}$--$M_{\rm halo}$ relation inferred from $M_*$ using abundance matching of $M_*$ with $M_{\rm halo}$ at $z \approx 0$ via the Universe Machine \citep{Behroozi2019MNRAS.488.3143B}. All symbols represent the same quantities as in Figure \ref{fig:MBH_Mhalo}, except that the \textit{entire} \citet{Terrazas_2017ApJ...844..170T} sample is shown, not just the subset without X-ray measurements from \citet{Gaspari2019}.
\label{fig:MBH-sigmav-Mstar}}
\end{figure*}

According to Figure \ref{fig:MBH_Mhalo}, black holes with $M_{\rm BH} \approx 10^8 \, M_\odot$ tend be found in halos close to $10^{13} \, M_\odot$ in mass, but in IllustrisTNG they reside in halos an order of magnitude less massive. Is it possible that the $M_{\rm halo}$ proxies shown in Figure \ref{fig:MBH_Mhalo} overestimate the mean halo masses of black holes with $M_{\rm BH} \approx 10^8 M_\odot$ by nearly an order of magnitude? That is the size of the adjustment needed to align the observations with the IllustrisTNG $M_{\rm BH}$--$M_{\rm halo}$ relation. To explore that possibility, we can consider what happens when halo masses are inferred from mass proxies other than X-ray temperature. 

\subsubsection{$M_{\rm BH}$ and $\sigma_v$}

The left panel of Figure \ref{fig:MBH-sigmav-Mstar} illustrates the $M_{\rm BH}$--$M_{\rm halo}$ relations obtained using $\sigma_v$ as a mass proxy. Points based on X-ray data have been removed, but all other symbols remain as they were in Figure \ref{fig:MBH_Mhalo}. The $M_{\rm BH}$--$\sigma_v$ relation has been transmuted into $M_{\rm BH}$--$M_{\rm halo}$ using equation (\ref{eq:Mhalo-sigmav}). It follows the power-law $M_{\rm BH}$--$M_{\rm halo}$ relation predicted by the baryon lifting hypothesis up to $\sigma_v \sim 240 \, {\rm km \, s^{-1}}$, at which $M_{\rm halo} \sim 10^{13.1} M_\odot$ and $M_{\rm BH} \sim 10^9 M_\odot$. Beyond there, the $M_{\rm BH}$--$\sigma_v$ relation becomes much steeper than $M_{\rm BH} \propto \sigma_v^5$. However, the observed $M_{\rm BH}$--$T_{\rm X}$ relations show no such break at the same location. Comparing with Figure \ref{fig:MBH_Mhalo} demonstrates that the steeper trend arises because $\sigma_v$ is no longer a good proxy for $M_{\rm halo}$.
This upturn in the $M_{\rm BH}$--$\sigma_v$ relation is well known \citep[e.g.,][]{McConnellMa_2013ApJ...764..184M,
Bogdan_2018ApJ...852..131B,SahuGraham_2019ApJ...887...10S}, and indicates that some physical process  \citep[such as the ``black hole feedback valve" outlined in][]{Voit_2020ApJ...899...70V} prevents the velocity dispersion of a halo's central galaxy from rising in proportion with the halo's maximum circular velocity once it reaches $\sigma_v \sim 240 \, {\rm km \, s^{-1}}$.

\subsubsection{$M_{\rm BH}$ and $M_*$}
\label{sec:Proxies-Mstar}

On the right side of Figure \ref{fig:MBH-sigmav-Mstar}, $M_{\rm halo}$ is inferred from $M_*$ via the abundance-matching fit of \citet{Behroozi2019MNRAS.488.3143B} at $z \approx 0$. The scatter in $M_{\rm BH}$ at fixed $M_*$ is impressively large, indicating that $M_*$ is a poor halo mass proxy for this purpose. Stellar bulge mass might be a better proxy for halo mass, given its tighter correlation with $M_{\rm BH}$ \citep[e.g.,][]{HaringRix_2004ApJ...604L..89H,Gultekin_2009ApJ...698..198G,KormendyHo2013ARAA..51..511K,McConnellMa_2013ApJ...764..184M,SavorgnanGraham_2016ApJ...817...21S}, but certain features of the $M_{\rm BH}$--$M_*$ relation suggest that $M_{\rm BH}$ may anticorrelate with $M_*$ at fixed halo mass. 

For example, consider just the red triangles (both filled and unfilled) representing quenched galaxies, which tend to be bulge-dominated. Nine such triangles near $M_* \sim 10^{10.7} M_\odot$ also have $M_{\rm BH} > 10^8 M_\odot$ and seem to be consistent with the IllustrisTNG $M_{\rm BH}$--$M_{\rm halo}$ relation. However, both Figure \ref{fig:MBH_Mhalo} and the left panel of Figure \ref{fig:MBH-sigmav-Mstar} show no data points in that region, because both $\sigma_v$ and $kT_{\rm X}$ for those galaxies indicate greater halo masses. The median temperature among those nine galaxies is $kT_{\rm X} \approx 0.3 \, {\rm keV}$, and the median velocity dispersion is $\sigma_v \approx 238 \, {\rm km \, s^{-1}}$, implying a median halo mass ($\sim 10^{13} M_\odot$) that places those same galaxies closer to the EAGLE $M_{\rm BH}$--$M_{\rm halo}$ relation. Systematic uncertainties among various sets of $M_{\rm halo}$ proxies might therefore explain why the apparent dispersion of $M_{\rm BH}$ at $M_{\rm halo} \sim 10^{12-12.5} M_\odot$ is so large in data sets that combine several different halo-mass proxies \citep[see e.g., Figure 1 of][]{Marasco_2021MNRAS.507.4274M}.

The anti-correlation between sSFR and $M_{\rm BH}$ found by \citet{Terrazas_2016ApJ...830L..12T,Terrazas_2017ApJ...844..170T} in their sample provides a clue as to why the scatter in $M_{\rm BH}$ at fixed $M_*$ is so large. At any given halo mass, stellar masses within the star-forming subset of galaxies are still increasing, while the stellar masses of the quiescent subset are not. It is therefore likely that some of the quiescent galaxies have stellar masses that are unusually small for their halo mass. Additionally, black hole masses in the quiescent population might be unusually large for their stellar mass, precisely because they have already experienced episodes of rapid black-hole growth that have lifted the halo's baryons and quenched star formation, resulting in a large dispersion in $M_{\rm BH}$ near $M_* \sim 10^{10.7} M_\odot$, where the quiescent and star-forming populations strongly overlap.

\subsubsection{Feedback and $T_{\rm CGM}$}

Another possibility to assess is that the \citet{Gaspari2019} galaxies with $kT_{\rm CGM} \sim 0.3 \, {\rm keV}$ are indicating halo masses that are approximately an order of magnitude too large. If $M_{\rm halo}$ is indeed overestimated by that much, then correcting for the overestimate would place those galaxies on the IllustrisTNG relation, with $M_{\rm BH} \sim 10^8 M_\odot$ corresponding to $M_{\rm halo} \sim 10^{12} M_\odot$. For example, such an overestimate might happen if kinetic feedback produces temperature fluctuations several times greater than what $T_{\rm CGM}$ would be in hydrostatic equilibrium. 

\begin{figure}[t]
\begin{center}
\includegraphics[width=3.4in,trim=0.2in 0.2in 0.0in 0.0in]{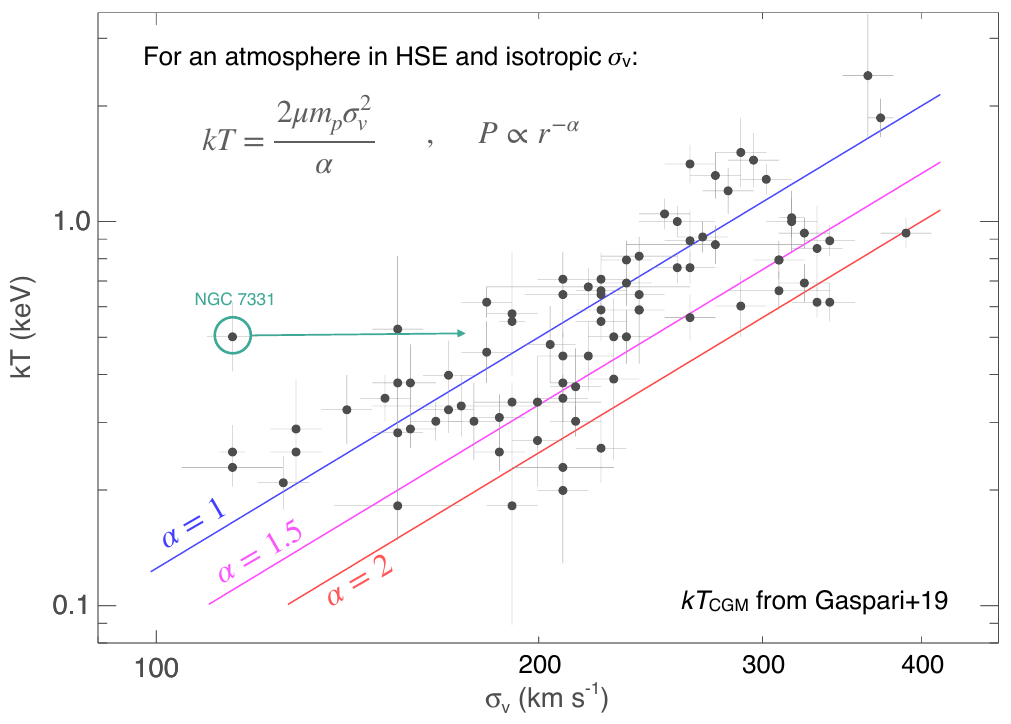}
\end{center}
\caption{Relationship between $\sigma_v$ and $kT_{\rm CGM}$ in the \citet{Gaspari2019} galaxy sample. Colored lines indicate hydrostatic temperatures corresponding to $\sigma_v$ for $\alpha \equiv | d \ln P / d \ln r |$ equal to 1, 1.5, and 2, as labeled. These values of $\alpha$ are representative of the range observed among massive elliptical galaxies. There is no obvious departure from those relations at low $\sigma_v$ that would indicate a temperature enhancement produced by black-hole feedback. The only outlier is NGC~7331, which has an observed rotation speed $v_{\rm c} \approx 250 \, {\rm km \, s^{-1}}$ \citep{Bottema_1999A&A...348...77B}, indicating that $\sigma_v$ does not reflect its halo mass. A green arrow shows where NGC~7331 ends up if $v_{\rm c}/\sqrt{2}$ is used instead of $\sigma_v$.
\vspace*{2em}
\label{fig:Gaspari19_sigmav_kt}}
\end{figure}

\citet{Truong_2021MNRAS.501.2210T} have performed mock X-ray observations of IllustrisTNG galaxies showing that the TNG feedback mechanism does produce biases in apparent temperature large enough to account for the apparent offset in halo mass. However, CGM temperatures in the \citet{Gaspari2019} sample show no evidence for such large departures from hydrostatic equilibrium. Figure~\ref{fig:Gaspari19_sigmav_kt} presents the relationship between $\sigma_v$ and $kT_{\rm CGM}$ in that sample, along with three lines representing the hydrostatic relation
\begin{equation}
    kT = \frac {2 \mu m_p \sigma_v^2} {\alpha} 
            \; \; \; \; \; , \; \; \; \; \; 
    \alpha \equiv \left| \frac { d \ln P} {d \ln r} \right| 
    \; \; ,
\end{equation}
for $\alpha = 1$, 1.5, and 2, given an isotropic velocity dispersion. Those values of $\alpha$ are representative for this sample and account for the spread in $kT_{\rm CGM}$ at fixed $\sigma_v$. If there were a feedback-induced departure from the hydrostatic relations below 0.5~keV, one would expect to see an excess of galaxies above the $\alpha = 1$ line at low $\sigma_v$, but there is just one outlier there. It is NGC~7331, which has $\sigma_v = 115 \, {\rm km \, s^{-1}}$ near its center but $v_c \approx 250 \, {\rm km \, s^{-1}}$ at 30~kpc \citep{Bottema_1999A&A...348...77B}, indicating a greater halo mass than its central stellar velocity dispersion implies. That circular velocity is equivalent to $\sigma_v \approx 180 \, {\rm km \, s^{-1}}$, making the CGM temperature of NGC~7331 consistent with hydrostatic equilibrium at $\alpha \approx 1$.

\subsubsection{A Closer Look at TNG}

We therefore conclude that the IllustrisTNG $M_{\rm BH}$--$M_{\rm halo}$ relation is in strong tension with the available observational constraints. Those simulations consequently seem to be inconsistent with the proposed three-way link between black hole growth, baryon lifting, and quenching of star formation, but they are not. The rest of the paper looks more closely at IllustrisTNG and shows that both black hole growth and quenching of star formation are indeed linked to baryon lifting, despite the anomalous $M_{\rm BH}$--$M_{\rm halo}$ relation. Sections \ref{sec:lifting} and \ref{sec:BHmode} outline how baryon lifting in IllustrisTNG is linked to star formation and black hole growth. Sections \ref{sec:price} and \ref{sec:priceisright} explain why the IllustrisTNG $M_{\rm BH}$--$M_{\rm halo}$ relation is anomalous and discuss what might be done to improve it.

\section{Lifting and Quenching}
\label{sec:lifting}

\begin{figure*}[th]
\begin{center}
\includegraphics[width=6.9in,trim=0.2in 0.0in 0.0in 0.0in]{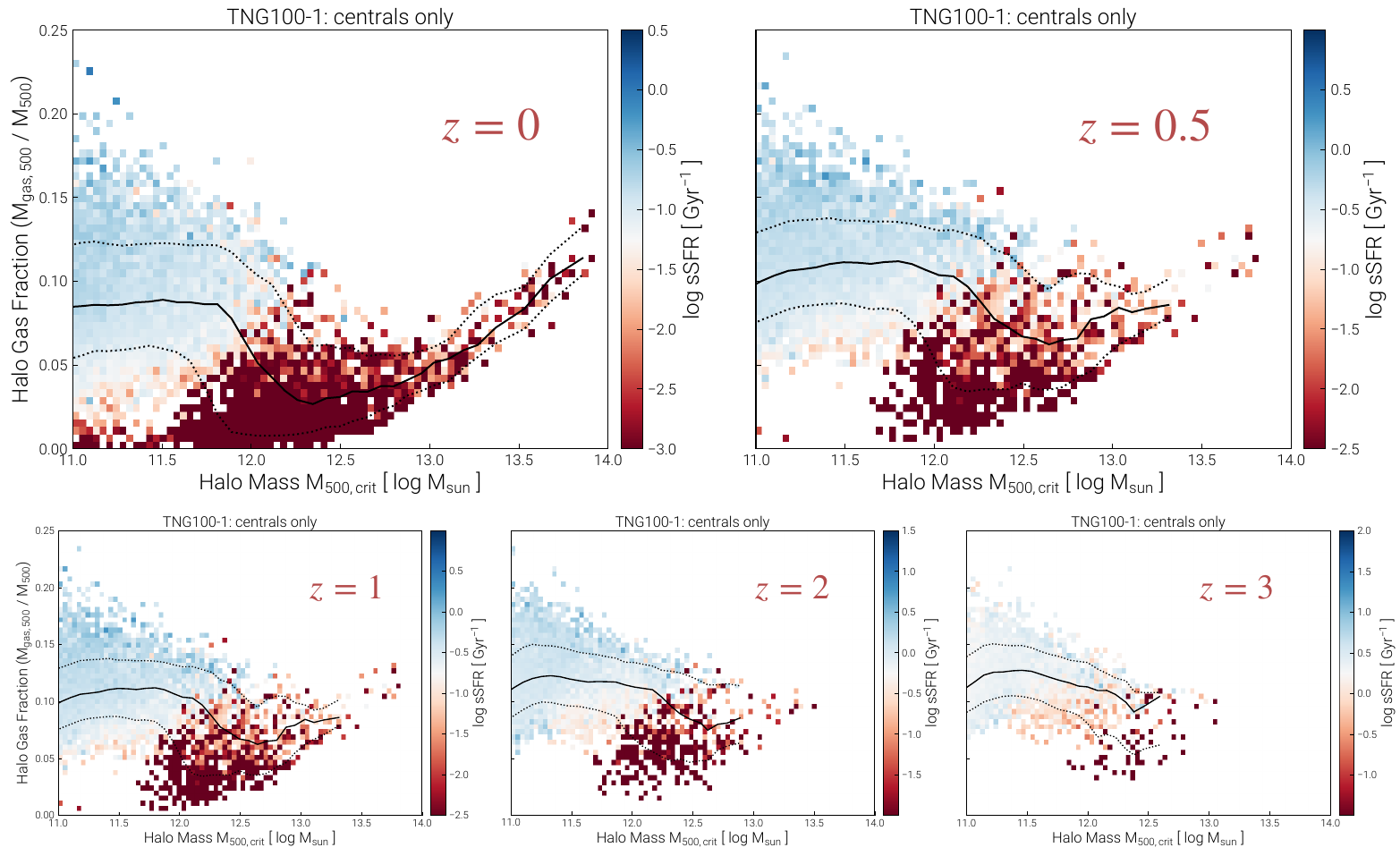}
\end{center}
\caption{Dependence of specific star formation rate (sSFR) on halo gas fraction and halo mass ($M_{\rm 500c}$) across the redshift range $0 \leq z \leq 3$ in the TNG100 simulation. Solid black lines show the median halo gas fraction at each halo mass, and dotted lines show the 10th and 90th percentiles. Colored squares show the typical sSFR associated with each combination of halo mass and halo gas fraction. Red squares representing low sSFR are systematically associated with lower halo gas fraction, indicating that star-formation quenching in IllustrisTNG galaxies is linked to feedback that lifts a halo's baryons to greater altitudes. (The halo gas fractions plotted here and in Figure \ref{fig:MBH-Ekin-fgas} include gas at all temperatures.)
\vspace*{2em}
\label{fig:M500-fgas-sSFR}}
\end{figure*}

\begin{figure*}[th]
\begin{center}
\includegraphics[width=6.8in,trim=0.2in 0.0in 0.0in 0.0in]{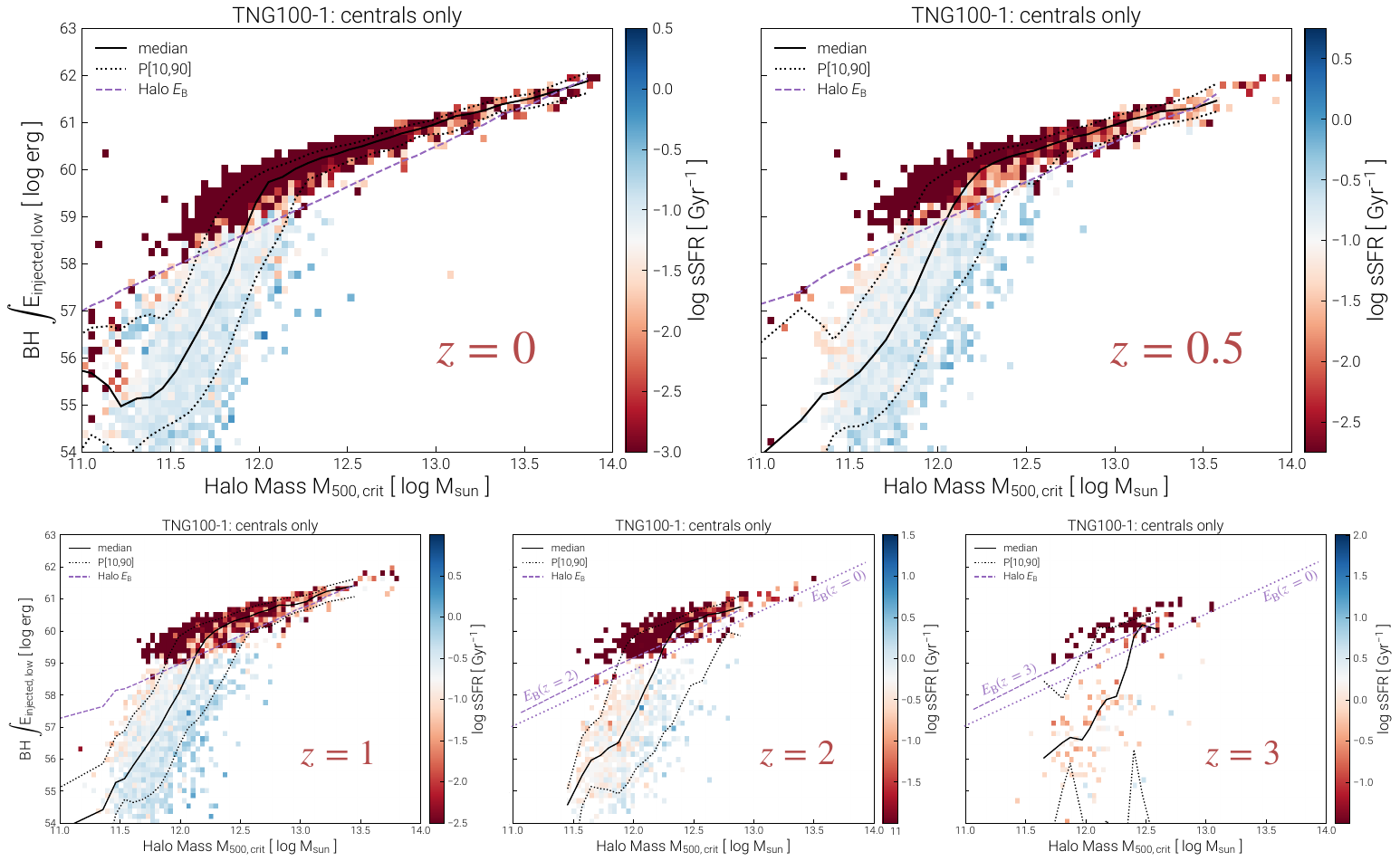}
\end{center}
\caption{Dependence of specific star formation rate (sSFR) on cumulative kinetic black hole  feedback ($E_{\rm kin} = \int E_{\rm injected, low}$) and halo mass ($M_{\rm 500c}$) across the redshift range $0 \leq z \leq 3$ in the TNG100 simulation. Solid black lines show the median amount of cumulative kinetic energy injection at each halo mass, and dotted lines show the 10th and 90th percentiles. Colored squares show the typical sSFR associated with each combination of halo mass and injected energy. Purple dashed lines show the characteristic scale of baryonic binding energy ($E_{\rm B}$) at each halo mass, and purple dotted lines in the lower right panels show the $M_{\rm 500c}$--$E_{\rm B}$ relation at $z=0$. At each halo mass and across all redshifts, the dark red squares indicating quenched star formation ($\rm{sSFR} \lesssim 10^{-2} \, {\rm Gyr}^{-1}$) are almost entirely above the dashed lines, and the lower edge of the quenched galaxy population tracks those lines. This correspondence implies that IllustrisTNG galaxies become quenched when kinetic feedback injects energy sufficient to lift the halo's baryons, thereby lowering the gas pressure and increasing the cooling time of hot gas near the central black hole. 
\vspace*{2em}
\label{fig:M500-Ekin-sSFR}}
\end{figure*}

Previous work has already established that baryon lifting coincides with star-formation quenching in both the IllustrisTNG and EAGLE cosmological simulations \citep{Bower_2017MNRAS.465...32B,Davies_2019MNRAS.485.3783D,Davies_2020MNRAS.491.4462D,Oppenheimer_2020MNRAS.491.2939O,Terrazas2020MNRAS.493.1888T,Zinger_2020MNRAS.499..768Z,Piotrowska_2022MNRAS.512.1052P}. Figure~\ref{fig:M500-fgas-sSFR} illustrates one of the key findings: central galaxies with quenched star formation in the TNG100 simulation have less halo gas than galaxies with active star formation.\footnote{
Figures \ref{fig:M500-fgas-sSFR} through \ref{fig:MBH-Ekin-fgas} come from the IllustrisTNG plotting tool at \url{https://www.tng-project.org/data/groupcat/} thanks to  \citet{Nelson_2019_PublicTNG_ComAC...6....2N}.} 
The figure plots the halo gas mass fraction ($f_{\rm gas} \equiv M_{\rm gas,500}/M_{\rm 500c}$) as a function of $M_{\rm 500c}$. Colors indicate the median sSFR at each combination of $f_{\rm gas}$ and $M_{\rm 500c}$. Red squares representing suppressed star formation are prevalent among halos of mass $M_{\rm 500c} \gtrsim 10^{12.5} \, M_\odot$ across the redshift range $0 \leq z \leq 3$. Among lower-mass halos, blue squares representing active star formation correspond to larger halo gas fractions than the red squares representing suppressed star formation. Galaxies in the EAGLE simulation follow the same qualitative trend \citep{Davies_2019MNRAS.485.3783D}, but $f_{\rm gas}$ in EAGLE is generally a factor of $\sim 3$ smaller at $M_{\rm halo} \lesssim 10^{12} \, M_\odot$ than in IllustrisTNG \citep{Davies_2020MNRAS.491.4462D}. Baryon lifting in low-mass halos must therefore proceed somewhat differently in the two simulation environments.

Figure~\ref{fig:M500-Ekin-sSFR} shows that star-formation rates in IllustrisTNG are also closely related to the central black hole's cumulative kinetic energy input ($E_{\rm kin}$), which includes the kinetic energy released by smaller black holes that have merged with the central one. A purple dashed line in each panel represents the quantity
\begin{equation}
    E_{\rm B} \equiv 
        \frac {3} {5} 
        \frac {G M_{\rm 200c}^2} {r_{\rm 200}} 
          f_{\rm b} 
         \; \; ,
\end{equation}
which is an estimate of the initial binding energy of the halo's baryons, for the cosmic mean baryon fraction $f_{\rm b}$ \citep{Nelson_2018_OVI}. It corresponds to a uniform sphere and should not be considered exact. But notably, the transition to highly suppressed star formation (dark red squares) lies close to that line across the redshift range $0 \leq z \leq 3$, indicating that star formation becomes quenched when the kinetic energy input associated with black hole accretion exceeds the amount of energy required to lift the circumgalactic gas.

\citet{Terrazas2020MNRAS.493.1888T} presented similar results. Their Figure~4 shows that the sSFR of an IllustrisTNG galaxy starts to decline when $E_{\rm kin}$ exceeds the gravitational binding energy of gaseous baryons currently within the galaxy ($E_{\rm bind,gal}$) and declines much more rapidly once $E_{\rm kin} \gtrsim 10 \, E_{\rm bind,gal}$. Most of the quenched galaxies end up with $E_{\rm kin} \gg 100 \, E_{\rm bind,gal}$. Also, Figure~6 in \citet{Terrazas2020MNRAS.493.1888T} shows that $E_{\rm kin}$ among the quenched galaxies is typically an order of magnitude greater than the gravitational binding energy of the gaseous baryons remaining within the halo ($E_{\rm bind,halo}$). However, neither $E_{\rm bind,gal}$ nor $E_{\rm bind,halo}$ scales linearly with $E_{\rm B}$ because their values decline precipitously as black hole feedback starts to lift baryons out of both the galaxy and the halo that contains it. 

Figure~\ref{fig:M500-Ekin-sSFR} of this paper is therefore complementary to the figures in \citet{Terrazas2020MNRAS.493.1888T}, because it compares $E_{\rm kin}$ to an atmospheric binding energy scale ($E_{\rm B}$) that remains steady while feedback rapidly acts to lift the halo's baryons. The value of $E_{\rm B}$ at fixed $M_{\rm 500c}$ declines slowly with time because the specific binding energy of a cosmological halo\footnote{Bounded by a radius $\propto M_{\rm halo}^{1/3} \rho_{\rm cr}^{-1/3}$} is $\propto M_{\rm halo}^{2/3} \rho_{\rm cr}^{1/3}$  and the cosmological critical density $\rho_{\rm cr}$ declines as the universe ages. Dotted purple lines in Figure \ref{fig:M500-Ekin-sSFR} show how much greater $E_{\rm B}(M_{\rm 500c})$ is at $z \geq 2$ than at $z=0$. The lower envelope of the dark red squares representing quenching is correspondingly at greater $E_{\rm B}$.

Interestingly, the distribution of $E_{\rm kin}$ at fixed $M_{\rm 500c}$ among IllustrisTNG halos with quenched central galaxies becomes narrower as halo mass increases. Meanwhile, the median value of $E_{\rm kin}$ at fixed $M_{\rm 500c}$ converges toward $E_{\rm B}$, becoming nearly equal to it as halo mass approaches $\sim 10^{14} \, M_\odot$. In halos that are even more massive, $E_{\rm B}$ exceeds $E_{\rm kin}$.  This outcome is qualitatively consistent with the observed rise in $f_{\rm gas}$ as halo masses go from $\sim 10^{13.5} M_\odot$ to $\sim 10^{14.5} M_\odot$ \citep[e.g.,][]{Pratt_2009_REXCESS_LX-T, Sun+09,Lovisari_2015A&A...573A.118L,
Eckert_2021Univ....7..142E}.

\citet{Zinger_2020MNRAS.499..768Z} have shown how black hole feedback in IllustrisTNG alters the central entropy and cooling time in massive halos as star formation shuts down. Early feedback is overwhelmingly thermal and relatively ineffective at quenching star formation. The transition to quiescence does not happen until kinetic feedback becomes significant. During that transition to kinetic feedback, the entropy\footnote{
As represented by $K = kT n^{-2/3}$, where $kT$ is the gas temperature in energy units and $n$ is the number density of gas particles} 
of the circumgalactic atmosphere rises above $\sim 10 \, {\rm keV \, cm^2}$ and its cooling time rises above $\sim 1$~Gyr. 

In halos of mass $\lesssim 10^{13.5} \, M_\odot$, baryon lifting is a necessary consequence of the transition to kinetic feedback, because those increased entropy levels and cooling times correspond to gas densities smaller than $f_{\rm b}$ times the total matter density. Making the transition happen therefore requires an energy input roughly equivalent to $E_{\rm B}$. Initially, suppression of star formation in IllustrisTNG may result from ``ejective" feedback that expels cool gas clouds from the galaxy, but long-term quiescence requires ``preventative" feedback that limits the galaxy's supply of cold gas by increasing the entropy and cooling time of the circumgalactic medium (CGM), which entails lifting of the entire atmosphere.

\section{Modes of Black Hole Growth}
\label{sec:BHmode}

\begin{figure*}[th]
\begin{center}
\includegraphics[width=6.8in,trim=0.2in 0.0in 0.0in 0.0in]{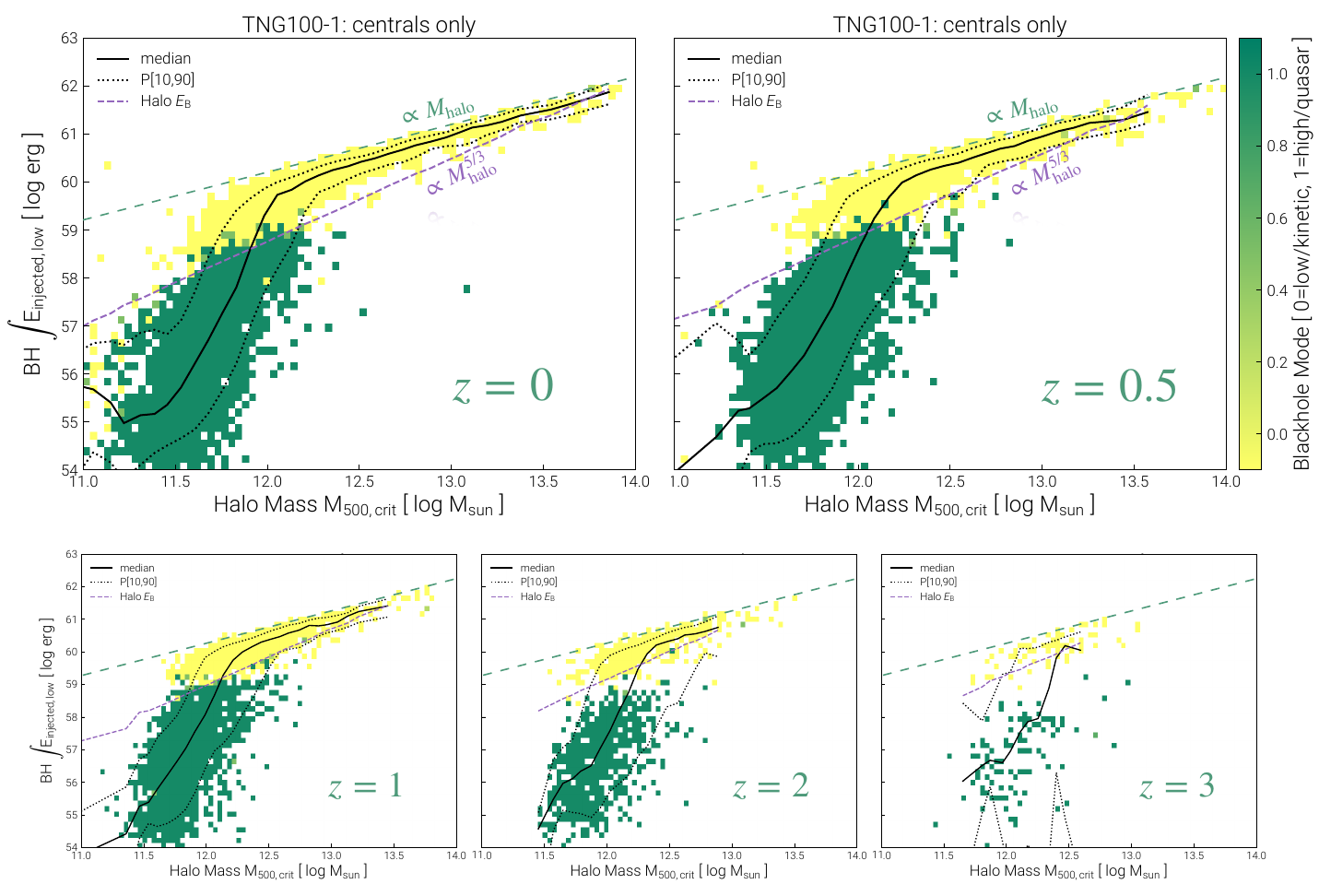}
\end{center}
\caption{Dependence of black hole feedback mode on cumulative kinetic black hole feedback ($E_{\rm kin} = \int E_{\rm injected, low}$) and halo mass ($M_{\rm 500c}$) across the redshift range $0 \leq z \leq 3$ in the TNG100 simulation. Solid black lines show the median amount of cumulative kinetic energy injection at each halo mass, and dotted lines show the 10th and 90th percentiles. Colored squares show the typical feedback mode associated with each combination of halo mass and injected energy: Green squares represent the thermal feedback mode associated with higher accretion rates, and yellow squares represent the kinetic feedback mode associated with lower accretion rates. Purple dashed lines approximately proportional to $M_{\rm halo}^{5/3}$ show the characteristic scale of baryonic binding energy ($E_{\rm B}$) at each halo mass. At each halo mass and across all redshifts, the yellow squares representing the kinetic feedback mode are almost entirely above that line, and the lower edge of the population in which kinetic feedback dominates tracks that line. The upper edge of that population tracks long-dashed green lines that are proportional to halo mass. Those bounds imply that the kinetic feedback mode in IllustrisTNG tunes itself to supply a total energy that is tied to the halo's baryonic binding energy in the mass range $10^{12} \, M_\odot \lesssim M_{\rm halo} \lesssim 10^{14} \, M_\odot$.
\vspace*{2em}
\label{fig:M500-Ekin-BHmode}}
\end{figure*}

Now we turn to the connection between black hole growth and suppression of star formation. The previous section showed that quenching in IllustrisTNG coincides with a cumulative kinetic energy input $E_{\rm kin}$ that exceeds the halo's baryonic binding energy scale. The transition to a quiescent state generally happens near $M_{\rm halo} \sim 10^{12} \, M_\odot$, at which $E_{\rm B} \sim 10^{59} \, {\rm erg}$. The corresponding amount of black hole mass growth is 
\begin{equation}
    \Delta M_{\rm BH} \: \gtrsim \: \frac {E_{\rm B}} {\epsilon_{\rm kin} c^2}
     \label{eq:DeltaMBH}
\end{equation}
in which $\epsilon_{\rm kin}$ is the conversion efficiency of accreted rest-mass energy into kinetic feedback energy and $c$ is the speed of light. In IllustrisTNG, this relationship results in  
\begin{equation}
    \Delta M_{\rm BH} \: \gtrsim \: 
       3 \times 10^5 \, M_\odot \left( \frac {E_{\rm B}} {10^{59} \, {\rm erg}} \right) 
       \label{eq:DeltaMBH_Illustris}
     \; \; ,
\end{equation}
given $\epsilon_{\rm kin} = 0.2$ \citep{Weinberger_2017MNRAS.465.3291W}. The total mass accumulated during kinetic-mode accretion is always subdominant compared to the mass accumulated during prior thermal-mode accretion because the efficiency factor assigned to kinetic-mode accretion is so large \citep{Weinberger_2018MNRAS.479.4056W}. Interpreting the $M_{\rm BH}$--$M_{\rm halo}$ relations emerging from IllustrisTNG therefore requires close attention to what governs the transition between feedback modes.

IllustrisTNG feedback modes depend on how the instantaneous accretion rate ($\dot{M}_{\rm BH}$) onto a halo's central black hole compares with the limiting Eddington rate 
\begin{equation}
  \dot{M}_{\rm Edd} = \frac {4 \pi G M_{\rm BH} m_p} {\epsilon_{\rm rad} \sigma_{\rm T} c}
   \; \; ,
\end{equation}
where $G$ is the gravitational constant, $m_p$ is the proton mass, $\sigma_{\rm T}$ is the Thomson electron scattering cross section, and $\epsilon_{\rm rad}$ is the conversion efficiency of accreted rest mass to radiative energy. In the fiducial IllustrisTNG model \citep{Weinberger_2017MNRAS.465.3291W}, black hole feedback is in thermal mode when
\begin{equation}
    \frac {\dot{M}_{\rm BH}} {\dot{M}_{\rm Edd}} \: > \: 
       \min \left[ 0.002 \left( 
                                \frac {M_{\rm BH}} 
                                      {10^8 \, M_\odot} \right)^2 , 0.1\right]
            \; \; .
            \label{eq:BHmode}
\end{equation}
Otherwise, the feedback mode is kinetic. The mass-dependent factor in equation (\ref{eq:BHmode}) favors kinetic feedback as $M_{\rm BH}$ rises above $\sim 10^8 \, M_\odot$. However, the thermal mode is still active when $\dot{M}_{\rm BH} > 0.1 \dot{M}_{\rm Edd}$, even if the black hole is very massive. 

Figure~\ref{fig:M500-Ekin-BHmode} shows the joint dependence of feedback mode on both $M_{\rm 500c}$ and $E_{\rm kin}$ in the TNG100 simulation. The axes are identical to Figure~\ref{fig:M500-Ekin-sSFR}. Comparing the two figures shows that kinetic feedback prevails among galaxies with quenched star formation. Just as in Figure~\ref{fig:M500-Ekin-sSFR}, galaxies undergo a transition as $E_{\rm kin}$ surpasses $E_{\rm B}$, switching to predominantly kinetic feedback. 

The twin transitions in both feedback mode and star formation behavior depend on two factors. First, a black hole's mass needs to approach $10^8 \, M_\odot$ for the TNG implementation of kinetic feedback to come into play. In the green regions of Figure~\ref{fig:M500-Ekin-BHmode}, where the thermal mode dominates, episodes of kinetic feedback must still sometimes occur, because $E_{\rm kin}$ is rising toward $E_{\rm B}$. Those kinetic feedback episodes become increasingly likely as $M_{\rm BH}$ grows, because of the relationship in equation (\ref{eq:BHmode}). Eventually the kinetic mode dominates, resulting in both star-formation quenching and baryon lifting. Second, thermal mode feedback becomes strongly disfavored as a transitioning galaxy loses its cold, dense clouds. The reason is that $\dot{M}_{\rm BH}$ in IllustrisTNG is taken to be the local Bondi accretion rate \citep{Bondi_1952MNRAS.112..195B}, which depends strongly on the specific entropy of accreting gas.\footnote{
The Bondi accretion rate is $\propto M_{\rm BH}^2 K^{-3/2}$.} 
Whenever the black hole is surrounded by the hot, high-entropy ambient gas characteristic of a quenched galaxy, accretion is slower, making the kinetic feedback more likely.

Previous analyses of quenching and feedback mode in IllustrisTNG have focused more closely on the role of $M_{\rm BH}$ than on multiphase gas and its role in black hole fueling. For example, \citet{Weinberger_2018MNRAS.479.4056W} showed that the median sSFR of IllustrisTNG galaxies dramatically drops as a direct result of kinetic feedback as $M_{\rm BH}$ rises above $\sim 10^{8.2} \, M_\odot$. \citet{Terrazas2020MNRAS.493.1888T} came to a similar conclusion and also showed that $f_{\rm gas}$ dramatically declines at the same black hole mass threshold.

\begin{figure*}[th]
\begin{center}
\includegraphics[width=6.8in,trim=0.2in 0.0in 0.0in 0.0in]{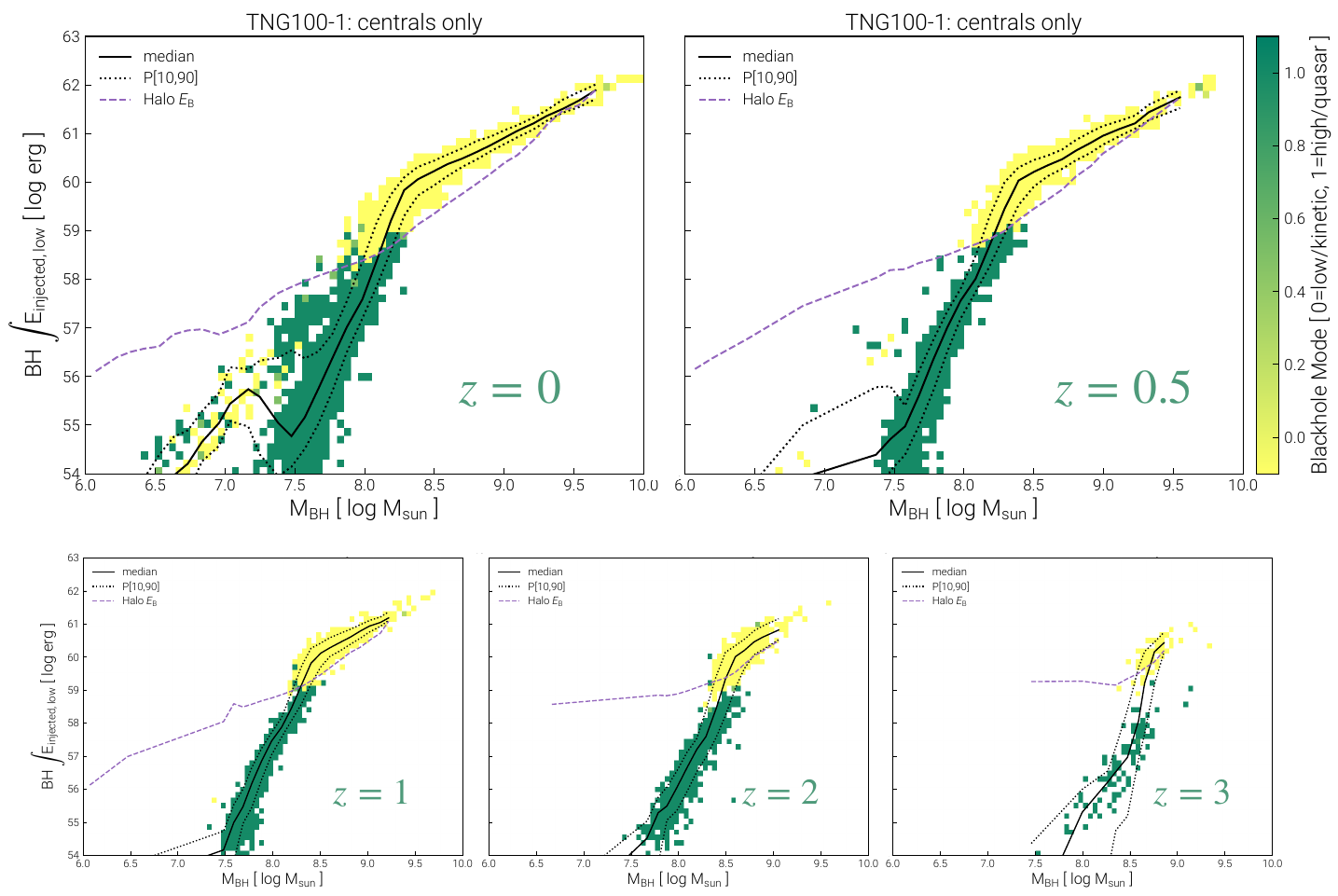}
\end{center}
\caption{Dependence of black hole feedback mode on cumulative kinetic black hole feedback ($E_{\rm kin} = \int E_{\rm injected, low}$) and black hole mass ($M_{\rm BH}$) across the redshift range $0 \leq z \leq 3$ in the TNG100 simulation. Solid black lines show the median amount of cumulative kinetic energy injection at each halo mass, and dotted lines show the 10th and 90th percentiles. Colored squares show the typical feedback mode associated with each combination of halo mass and injected energy: Green squares represent the thermal feedback mode associated with higher accretion rates, and yellow squares represent the kinetic feedback mode associated with lower accretion rates. Purple dashed lines show the characteristic scale of baryonic binding energy ($E_{\rm B}$) at each black hole mass. The transition to dominant kinetic feedback occurs as the cumulative kinetic energy input surpasses the halo's baryonic binding energy and happens at lower black hole masses within less massive halos. Consequently, cumulative kinetic feedback input \textit{prior to the transition} is the primary cause of that transition.
\vspace*{2em}
\label{fig:MBH-Ekin-BHmode}}
\end{figure*}

Superficially, quenching of star formation in IllustrisTNG may seem to depend most strongly on the threshold value of $M_{\rm BH}$ marking the onset of kinetic feedback, but cumulative kinetic energy input ($E_{\rm kin}$) turns out to be even more critical \citep{Terrazas2020MNRAS.493.1888T}. Figure~\ref{fig:MBH-Ekin-BHmode} shows the joint dependence of black hole feedback mode on both $M_{\rm BH}$ and $E_{\rm kin}$. As in Figure~\ref{fig:M500-Ekin-BHmode}, the transition to kinetic mode depends most directly on how $E_{\rm kin}$ compares with $E_{\rm B}$. If a threshold in $M_{\rm BH}$ were more critical, then the boundary between the green and yellow regions would be vertical in Figure~\ref{fig:MBH-Ekin-BHmode}. Instead, the boundary is diagonal and closely coincides with the line marking $E_{\rm kin} = E_{\rm B}$.

Figure~\ref{fig:MBH-Ekin-BHmode} also shows that the transitional values of $M_{\rm BH}$ are larger in higher-redshift galaxies. Given how $\dot{M}_{\rm BH} / \dot{M}_{\rm Edd}$ determines the feedback mode, this redshift dependence indicates that the Bondi accretion rates onto the most massive black holes in IllustrisTNG are generally greater early in time than later in time, causing the thermal mode to dominate among black holes with masses approaching $10^{8.5} \, M_\odot$ at $z \approx 3$. At lower redshifts, the transitional value of $M_{\rm BH}$ clearly correlates with $E_{\rm kin}$ and does not exceed $M_{\rm BH} \approx 10^{8.3} \, M_\odot$ at $z = 0$ for any value of $E_{\rm kin}$. This decline with time in the maximum $M_{\rm BH}$ at which thermal mode feedback occurs implies that the typical specific entropy ($K \propto kT n^{-2/3}$) of gas near black holes of mass $M_{\rm BH} \sim 10^{8.3-8.5} \, M_\odot$ is lower at $z \sim 3$ than at $z \sim 0$, corresponding to greater pressure at a given gas temperature, presumably because of greater gas accretion rates onto those high-redshift galaxies.

\begin{figure*}[th]
\begin{center}
\includegraphics[width=6.6in,trim=0.2in 0.0in 0.0in 0.0in]{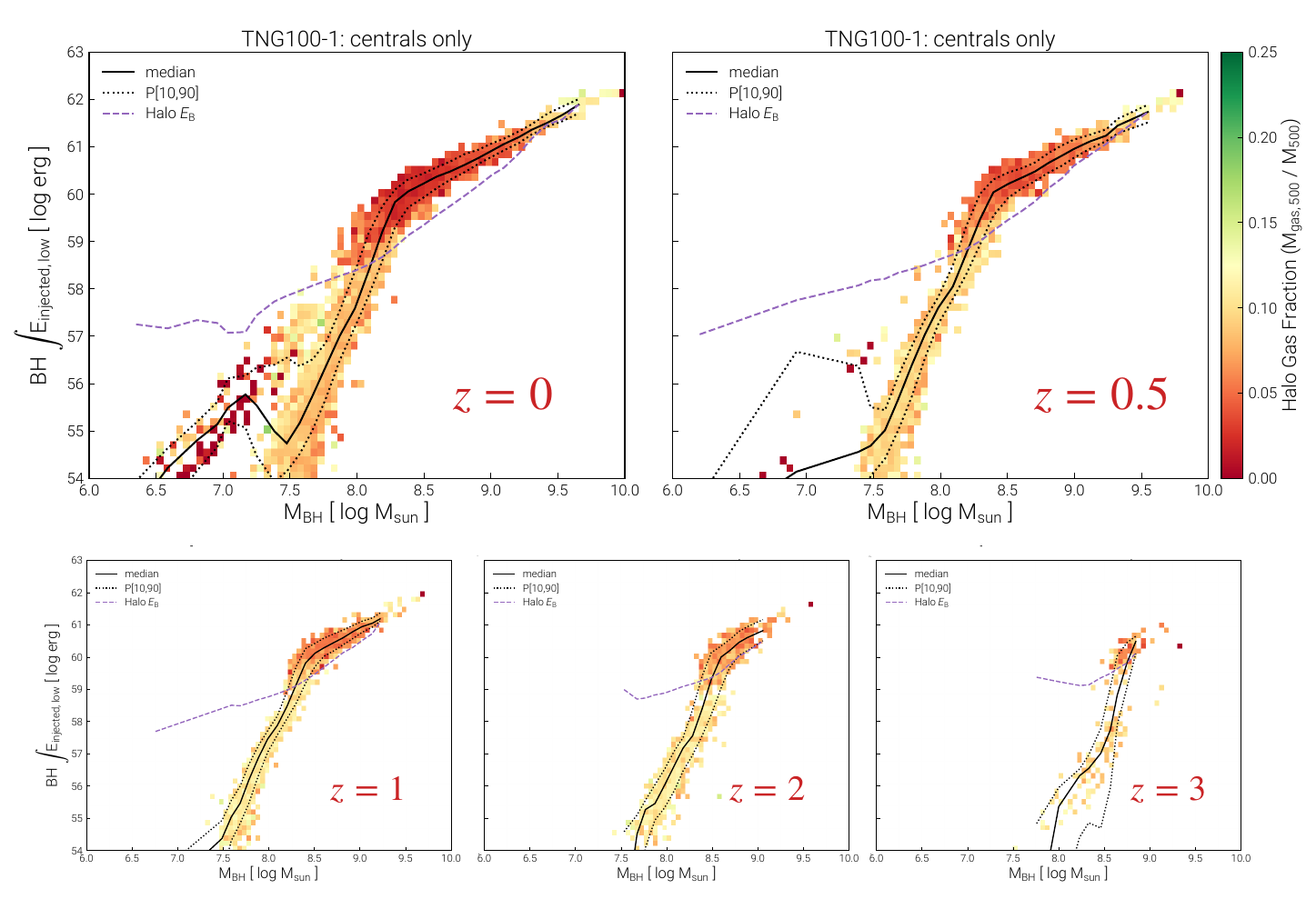}
\end{center}
\caption{Dependence of halo gas fraction on cumulative kinetic black hole feedback ($E_{\rm kin} = \int E_{\rm injected, low}$) and black hole mass ($M_{\rm BH}$) across the redshift range $0 \leq z \leq 3$ in the TNG100 simulation. Solid black lines show the median amount of cumulative kinetic energy injection at each halo mass, and dotted lines show the 10th and 90th percentiles. Colored squares show the halo gas fraction associated with each combination of halo mass and injected energy: Redder squares represent halo gas fractions substantially lower than the cosmic mean baryon fraction, indicating that feedback has lifted the halo's baryons. Purple dashed lines show the characteristic scale of baryonic binding energy ($E_{\rm B}$) at each black hole mass. The transition to low halo gas fractions occurs as the cumulative kinetic energy input surpasses the halo's baryonic binding energy and coincides with the transitions to both star-formation quenching and a shutdown in thermal mode feedback.
\vspace*{2em}
\label{fig:MBH-Ekin-fgas}}
\end{figure*}

The diagonal trend of each green region in Figure~\ref{fig:MBH-Ekin-BHmode}, sharply rising from lower left to upper right, indicates that episodes of kinetic feedback still sometimes occur among black holes below the transitional mass, but they must be rare. In the panels of Figure~\ref{fig:MBH-Ekin-BHmode}, some of the black holes near $M_{\rm BH} \sim 10^{7.5} \, M_\odot$ have managed to generate $E_{\rm kin} \gtrsim 10^{54} \, {\rm erg}$. As their cumulative kinetic output then grows to reach $\sim 10^{59} \, {\rm erg}$, those black holes accrete another $\sim 10^8 \, M_\odot$. Only $\sim 3 \times 10^5 \, M_\odot$ of that mass increase comes from accretion associated with kinetic feedback (see equation \ref{eq:DeltaMBH}), corresponding to $\lesssim 0.3$\% of the total.

Figure~\ref{fig:MBH-Ekin-fgas} confirms that star-formation quenching does indeed coincide with baryon lifting brought about by kinetic feedback. It illustrates how $f_{\rm gas}$ depends jointly on $M_{\rm BH}$ and $E_{\rm kin}$. In general, the red and dark orange points indicating large reductions in halo gas density lie above the purple dashed lines marking $E_{\rm kin} = E_{\rm B}$. Those same lines also mark the transition to a quenched state in  Figure \ref{fig:M500-Ekin-sSFR}. Except for the spurious tail of dark red squares at low $E_{\rm kin}$ and $M_{\rm BH}$ in the $z = 0$ panel of Figure~\ref{fig:MBH-Ekin-fgas} (to be discussed in \S \ref{sec:price}), there is no systematic dependence of $f_{\rm gas}$ on either $E_{\rm kin}$ or $M_{\rm BH}$ below the purple dashed lines. The lack of dependence on $M_{\rm BH}$ implies that thermal mode feedback does not contribute to baryon lifting, because cumulative thermal energy injection by the thermal mode is proportional to $M_{\rm BH}$. 

In contrast, reductions in halo gas content clearly depend on how $E_{\rm kin}$ compares to $E_{\rm B}$, with the greatest reductions in $f_{\rm gas}$ corresponding to $E_{\rm kin} \gg E_{\rm B}$ \citep[see also][]{Terrazas2020MNRAS.493.1888T}. At $M_{\rm BH} \approx 10^{8.3} \, M_\odot$ in the $z=0$ panel of Figure~\ref{fig:MBH-Ekin-fgas}, the median $E_{\rm kin}$ exceeds $E_{\rm B}$ by an order of magnitude, and that is where the reductions in $f_{\rm gas}$ are greatest. The excess of $E_{\rm kin}$ over $E_{\rm B}$ then declines with increasing $M_{\rm BH}$, until the two quantities are almost equal at $M_{\rm BH} \approx 10^{9.5} \, M_\odot$, where $E_{\rm B} \approx 10^{61.5} \, {\rm erg}$. In that part of Figure \ref{fig:MBH-Ekin-fgas} the squares indicating $f_{\rm gas}$ are typically yellow. Therefore, baryon lifting is minimal for $E_{\rm kin} \lesssim E_{\rm B}$ at both low and high halo masses.

Convergence of $f_{\rm gas}$ back toward the cosmic mean at high masses is consistent with the general trend observed among real galaxy groups and clusters and can be understood in terms of radiative cooling. Figures \ref{fig:M500-Ekin-sSFR} and \ref{fig:M500-Ekin-BHmode} show that $M_{\rm BH} \approx 10^{9.5} \, M_\odot$ corresponds to $M_{\rm 500c} \approx 10^{14} \, M_\odot$ at $z=0$ in the IllustrisTNG universe. In the real universe, halos with similar masses currently have X-ray luminosities $\sim 10^{44} \, {\rm erg \, s^{-1}}$, meaning that they can radiate $\sim 10^{61.5} \, {\rm erg}$ over the course of cosmic time, thereby converting a comparable amount of injected feedback energy into photons rather than into atmospheric gravitational potential energy. Consequently, black hole feedback in galaxy clusters ($M_{\rm halo} \sim 10^{14-15} \, M_\odot$) can self-regulate by balancing radiative losses, without much baryon lifting.

However, different simulations make strikingly different predictions for the radial distributions of baryons in and around massive halos \citep{Oppenheimer_2021Univ....7..209O,Sorini_2022MNRAS.516..883S}. Recently,  \citet{Ayromlou_2023MNRAS.524.5391A} compared the baryon distributions emerging from IllustrisTNG, EAGLE, and also the SIMBA simulation \citep{Dave_SIMBA_2019MNRAS.486.2827D}, finding that black hole feedback lifts a halo’s baryons least effectively in EAGLE and most effectively in SIMBA. The radial profiles of those baryon distributions are largely unconstrained by existing observations of halos below $10^{13.5} \, M_\odot$, but notably IllustrisTNG and EAGLE appear to overlap X-ray observations of massive groups more closely than SIMBA \citep{Oppenheimer_2021Univ....7..209O}.  

Differences among the simulations are to be expected, given how crude their black-hole feedback implementations remain. In EAGLE, that feedback is purely thermal, at a temperature chosen to minimize radiative losses \citep{Schaye_EAGLE_2015MNRAS.446..521S,CrainEAGLE+2015}. The kinetic feedback mode of IllustrisTNG injects energy through a series of randomly oriented impulses \citep{Weinberger_2017MNRAS.465.3291W}. SIMBA's kinetic black hole feedback is bipolar \citep{Dave_SIMBA_2019MNRAS.486.2827D}. None of those simulations reproduces the distinctive jet-lobe radio morphologies observed among massive halos with active feedback \citep{DonahueVoit2022PhR...973....1D}. Refinements of their feedback algorithms will benefit from paying close attention to observations of jets, X-ray cavities, and radio lobes, which reflect the jet power and zone of influence more directly than they reflect cumulative feedback energy. Nevertheless, cumulative black hole feedback energy in all of them suffices to lift the atmospheres of halos with $M_{\rm halo} \sim 10^{12.5}$--$10^{14} \,  M_\odot$, meaning that atmospheric lifting is linked to black hole growth in all such simulations, as long as feedback energy couples to the circumgalactic medium without significant radiative losses. 

\section{The Price of Feedback}
\label{sec:price}

Whether or not the masses of real black holes reflect the energy input required for quenching of star formation depends on the price of feedback. Assuming that baryon lifting is necessary for long-term quenching implies that a central black hole must inject an amount of energy at least as great as the halo's baryonic binding energy ($E_{\rm B}$) into the CGM. The injected energy comes at a ``price" of at least
\begin{equation}
    \Delta M_{\rm BH} \, \sim \, \frac {E_{\rm B}} {\epsilon_{\rm fb} c^2}
\end{equation}
in black hole mass growth that depends on the conversion efficiency $\epsilon_{\rm fb}$ of accreted rest-mass energy into feedback energy. If coupling of feedback energy to the CGM is highly inefficient, as happens during episodes of thermal mode feedback in IllustrisTNG, then the price can be much greater. 

\citet{DonahueVoit2022PhR...973....1D} show that the $M_{\rm BH}$--$T_{\rm CGM}$ relation obtained by \citet{Gaspari2019} is consistent with $M_{\rm BH} \approx 200 E_{\rm B} / c^2$. If this observed relationship does indeed reflect a connection between black hole mass and baryonic binding energy, then it implies a price of
\begin{equation}
     \Delta M_{\rm BH} \, \approx \, 10^7 \, M_\odot 
            \left( \frac {E_{\rm B}} {10^{59} \, {\rm erg}} \right)
            \label{eq:price}
\end{equation}
for star-formation quenching via baryon lifting. In the present-day universe, that price is
\begin{equation}
     \Delta M_{\rm BH} \, \approx \, 10^{8.3} \, M_\odot 
            \left( \frac {M_{\rm halo}} {10^{13} \, M_\odot} \right)^{1.6}
\end{equation}
when written in terms of halo mass, for halos in the mass range $10^{12.5} M_\odot \lesssim M_{\rm halo} \lesssim 10^{14} \, M_\odot$.

The simulations of \citet{BoothSchaye_2010MNRAS.405L...1B} produced a similar relationship using a feedback efficiency factor $\epsilon_{\rm fb} = 0.015$ and obtained black hole masses a factor of $\sim 2$ smaller at fixed $M_{\rm halo}$. The motivation for that choice of $\epsilon_{\rm fb}$ was to reproduce both the $M_{\rm BH}$--$M_*$ and $M_{\rm BH}$--$\sigma_v$ relations observed at $z \approx 0$ \citep{BoothSchaye_2009MNRAS.398...53B}. Feedback from black holes in their simulations (and the EAGLE simulations that ensued) is purely thermal but episodic and is released in pulses great enough to limit radiative losses of the injected feedback energy. It therefore couples far more efficiently with the CGM than the thermal mode feedback in IllustrisTNG.

The price of star-formation quenching in IllustrisTNG is considerably greater than the one in equation (\ref{eq:price}) because it contains both a fixed cost and a marginal cost. Quenching doesn't happen in a halo's central galaxy until the kinetic feedback mode introduces a cumulative energy comparable to $E_{\rm B}$. Its black hole mass must therefore exceed $\sim 10^8 \, M_\odot$, so that the condition in equation (\ref{eq:BHmode}) allows the kinetic mode to prevail. That is the fixed cost, and it establishes a ratio $M_{\rm BH}/M_{\rm halo} \sim 10^{-4}$ at the time of quenching in halos of mass $\sim 10^{12} \, M_\odot$. In comparison, the marginal cost of the kinetic feedback that \textit{actually} quenches star formation is miniscule, amounting to $3 \times 10^5 \, M_\odot$ for every $10^{59} \, {\rm erg}$ of energy injection (see equation \ref{eq:DeltaMBH_Illustris}). The total black-hole mass price for quenching in IllustrisTNG
\begin{equation}
    \Delta M_{\rm BH} \: \sim \: 
        10^{8} M_\odot 
        + 10^{5.5} M_\odot \left( \frac {M_{\rm halo}} 
                        {10^{12} M_\odot} \right)^{5/3}
\end{equation}
therefore depends almost entirely on the pivot mass in equation (\ref{eq:BHmode}) and can be lowered by reducing that pivot mass (see \citealt{Terrazas2020MNRAS.493.1888T} and \citealt{Truong_2021MNRAS.501.2210T}).

Once the cost to activate the kinetic mode has been paid, hierarchical merging ensures that the majority of a black hole's mass in IllustrisTNG still comes from thermal-mode accretion \citep{Weinberger_2018MNRAS.479.4056W}. Figure \ref{fig:M500-Ekin-BHmode} shows that kinetic feedback injects $\sim 10^{62} \, {\rm erg}$ during the history of a $\sim 10^{14} \, M_\odot$ halo, at a black hole mass cost of $\sim 3 \times 10^8 \, M_\odot$. Meanwhile, the central black hole's mass approaches $\sim 10^{10} M_\odot$ by consuming smaller black holes that grew to contain a fraction $\sim 10^{-4}$ of their halo's mass during the time of quenching.

Another consequence of hierarchical merging in IllustrisTNG is a mass-dependent upper limit on the value of $E_{\rm kin}$. Dashed green lines in Figure \ref{fig:M500-Ekin-BHmode} show that $E_{\rm kin}$ remains $\lesssim 10^{60} \, {\rm erg} \: ( M_{\rm halo} / 10^{12} \, M_\odot)$ as halo masses increase toward $\sim 10^{14} \, M_\odot$. In IllustrisTNG, black hole mergers preserve the sum of cumulative kinetic energy injection, and so $E_{\rm kin}$ reflects the entire history of kinetic energy injection associated with a particular halo. The upper edge of the relation between $E_{\rm kin}$ and halo mass therefore reflects the kinetic energy requirements for quenching at $M_{\rm halo} \sim 10^{12-12.5} \, M_\odot$. 

\begin{figure*}[th]
\begin{center}
\includegraphics[width=6.6in,trim=0.2in 0.0in 0.0in 0.0in]{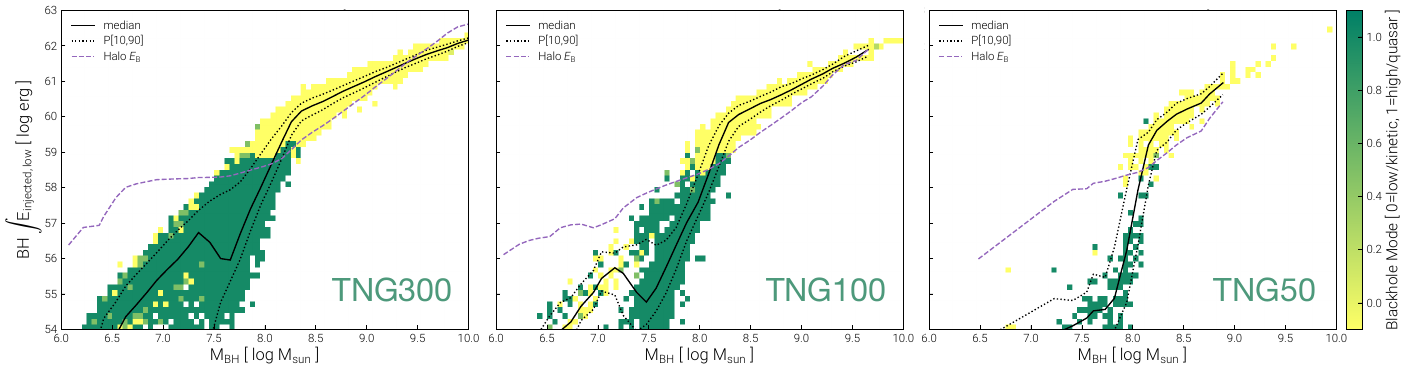}
\end{center}
\caption{Dependence of black hole feedback mode on spatial resolution. Each panel is a version of the upper left ($z=0$) panel of Figure~\ref{fig:MBH-Ekin-BHmode}, with TNG300 at left, TNG100 in the center, and TNG50 at right. Spatial resolution improves from left to right, and the total number of halos declines. The tail of points toward low $M_{\rm BH}$ and low $E_{\rm kin}$ in the left two panels is absent from the right panel, indicating that the tail results from insufficient spatial resolution. 
\label{fig:BHmode_resolution}}
\end{figure*}

Before we consider how the IllustrisTNG feedback parameters might be adjusted to bring the simulated $M_{\rm BH}$--$M_{\rm halo}$ relation into better agreement with observations, it is worth noting that the history of kinetic feedback immediately preceding star-formation quenching depends somewhat on numerical resolution. Figure~\ref{fig:BHmode_resolution} shows the dependence of feedback mode on both $M_{\rm BH}$ and $E_{\rm kin}$ in the TNG300, TNG100, and TNG50 simulations, proceeding toward finer spatial resolution from left to right. In the two lower resolution simulations, there is a tail of points at low $M_{\rm BH}$ and $E_{\rm kin}$ that is not present in TNG50. Also, the diagonal climb of the green region to the transition at $E_{\rm kin} \approx E_{\rm B}$ is steepest in TNG50. Apparently, the incidence of kinetic feedback episodes while $M_{\rm BH} < 10^8 \, M_\odot$ is smallest in TNG50, implying that improvements in spatial resolution raise the probability that there will be some low-entropy clouds, capable of fueling large Bondi accretion rates, close to the black hole. The points with low $f_{\rm gas}$ in the $z = 0$ tail of Figure~\ref{fig:MBH-Ekin-fgas} are therefore likely to be spurious results of insufficient spatial resolution near the central black hole. 

\section{What Price is Right?}
\label{sec:priceisright}

The IllustrisTNG feedback model depends on several parameters that were tuned to optimize agreement with observations of the stellar populations of galaxies \citep{Weinberger_2017MNRAS.465.3291W,Weinberger_2018MNRAS.479.4056W}. Adjustments of some of those parameters could potentially improve agreement with observational constraints on the $M_{\rm BH}$--$M_{\rm halo}$ relation. However, care must be taken not to degrade many other aspects of IllustrisTNG that agree with observations of galaxy evolution. 

The analyses in \S \ref{sec:BHmode} and \S \ref{sec:price} imply that the two most critical feedback parameters determining the $M_{\rm BH}$--$M_{\rm halo}$ relation in IllustrisTNG are $\epsilon_{\rm kin}$ and the $10^8 \, M_\odot$ pivot mass in equation (\ref{eq:BHmode}). Reduction of $\epsilon_{\rm kin}$ would appear necessary for better agreement with observational constraints, as its fiducial value ($\epsilon_{\rm kin} = 0.2$) results in a black hole mass price for baryon lifting at least an order of magnitude smaller than the one that \citet{DonahueVoit2022PhR...973....1D} infer from the $M_{\rm BH}$--$T_{\rm CGM}$ correlation presented by \citet{Gaspari2019}. The EAGLE simulations, which adopt the \citet{BoothSchaye_2010MNRAS.405L...1B} feedback efficiency, indicate that $\epsilon_{\rm kin} \approx 0.015$ might yield an $M_{\rm BH}$--$M_{\rm halo}$ relation in better alignment with observations. Equation (\ref{eq:price}) implies a lower limit of $\epsilon_{\rm kin} \gtrsim 0.005$, because further reduction would result in a black hole mass price for baryon lifting that exceeds observations.

More importantly, the pivot mass for switching to kinetic feedback in IllustrisTNG results in a black hole mass price prior to quenching that appears to be an order of magnitude larger than observations indicate. For example, the 24 star-forming galaxies in the \citet{Terrazas_2017ApJ...844..170T} sample belonging to the $10^{10.5} \, M_\odot$--$10^{11} \, M_\odot$ range of stellar mass have a median black hole mass $M_{\rm BH} = 10^{7.15} \, M_\odot$. The standard deviation around that median is 0.19 dex, and none of those galaxies has $M_{\rm BH} > 10^8 \, M_\odot$ (see Figure~\ref{fig:MBH-sigmav-Mstar}).

Reducing the pivot mass in equation (\ref{eq:BHmode}) would lower the maximum black-hole masses in star-forming IllustrisTNG galaxies. \citet{Truong_2021MNRAS.501.2210T} have explored the consequences of a reduction to $M_{\rm piv} = 10^{6.4} \, M_\odot$. That change results in a nearly linear $M_{\rm BH}$--$M_{\rm halo}$ relationship close to $M_{\rm BH} \approx 10^{-5} \, M_{\rm halo}$ for $10^{11.5} \, M_\odot \lesssim M_{\rm halo} \lesssim 10^{13.5} \, M_\odot$. It therefore improves agreement with the \citet{Terrazas_2017ApJ...844..170T} sample at the low-mass end but produces black hole masses that fall short of observations at the high-mass end.

Simultaneously implementing $\epsilon_{\rm kin} \approx 0.015$ and $M_{\rm piv} \approx 10^7 \, M_\odot$ in IllustrisTNG could potentially result in black hole masses that agree with $M_{\rm BH}$--$M_{\rm halo}$ observations at both the low- and high-mass ends. However, a large potential downside could be a reduction in both the halo mass and stellar mass at which star-formation quenching sets in, once black hole feedback becomes primarily kinetic. For example, $M_{\rm BH} \approx 10^7 \, M_\odot$ in the IllustrisTNG simulations corresponds to $M_{\rm halo} \sim 10^{11.3} \, M_\odot$ and $M_* \sim 10^{9.3} \, M_\odot$, both of which are significantly smaller than the observed quenching scales for central galaxies.

Another conceivable modification to the black hole feedback algorithm would be a transition from thermal to kinetic feedback that is not a step function of $\dot{M}_{\rm BH} / \dot{M}_{\rm Edd}$. In the current incarnation of IllustrisTNG, feedback during periods when $\dot{M}_{\rm BH} \gtrsim 0.1 \, \dot{M}_{\rm Edd}$ is entirely thermal, even though many quasars are known to have powerful winds and jets. Adding a kinetic feedback channel to the ``quasar" mode could qualitatively change how that feedback mode interacts with the surrounding atmosphere, even if the proportion of feedback energy in kinetic form is relatively small \citep[see, e.g.,][]{Meece_2017ApJ...841..133M}.

\section{Summary}
\label{sec:summary}


Observations gathered over the last couple of decades have long suggested that the masses of supermassive black holes are linked to the masses of the cosmological halos in which they reside \citep[e.g.,][]{Ferrarese_2002ApJ...578...90F,Bandara_2009ApJ...704.1135B,Marasco_2021MNRAS.507.4274M}. Those observations have repeatedly indicated a nearly linear relationship between $M_{\rm BH}$ and the binding energy of the halo's baryons ($E_{\rm B}$), in alignment with models of self-regulated black hole growth \citep[e.g.,][]{SilkRees1998AA...331L...1S,
Haehnelt_1998MNRAS.300..817H}. Among identically structured halos, the expected relationship would be $M_{\rm BH} \propto M_{\rm halo}^{5/3}$, but \citet{BoothSchaye_2010MNRAS.405L...1B} found a slightly shallower relationship ($M_{\rm BH} \propto M_{\rm halo}^{1.55}$) among cosmological halos with a more realistic dependence of halo concentration on halo mass.

X-ray analyses of the $M_{\rm BH}$--$T_{\rm X}$ relation \citep[e.g.,][]{Bogdan_2018ApJ...852..131B,Lakhchuara_2019MNRAS.488L.134L,Gaspari2019} have recently provided additional insights, because $T_{\rm X}$ supplies the most reliable estimates of $M_{\rm halo}$ for nearby galaxies with dynamical $M_{\rm BH}$ measurements. Notably, \citet{Gaspari2019} found that $M_{\rm BH}$ correlates more closely with circumgalactic gas temperature ($T_{\rm CGM}$) than with any other observable galactic or halo property. Figure~\ref{fig:MBH-TCGM-Thalo} shows that the $M_{\rm BH}$--$M_{\rm halo}$ relation found by converting $T_{\rm CGM}$ to $M_{\rm halo}$ using an observational $M_{\rm halo}$--$T_{\rm X}$ relation gives $M_{\rm BH} \propto M_{\rm halo}^{1.6}$, in excellent alignment with earlier constraints. It also extends that power-law $M_{\rm BH}$--$M_{\rm halo}$ relationship up to $M_{\rm halo} \sim 10^{14} M_\odot$, implying that the masses of black holes in galaxy groups reflect the energy input needed to lift their baryons \citep[see also][]{DonahueVoit2022PhR...973....1D}.

However, the $M_{\rm BH}$--$M_{\rm halo}$ relations emerging from cosmological numerical simulations are not as well aligned with the observational constraints (Figure \ref{fig:MBH_Mhalo}). Central black hole masses in EAGLE are close to the observational constraints for $M_{\rm halo} \sim 10^{11.5-13.5} M_\odot$ but underpredict $M_{\rm BH}$ in more massive halos. IllustrisTNG, on the other hand, produces black hole masses at $M_{\rm halo} \sim 10^{14} M_\odot$ in apparent agreement with observations but overpredicts $M_{\rm BH}$ at $M_{\rm halo} \sim 10^{12} M_\odot$. That happens because the $M_{\rm BH}$--$M_{\rm halo}$ relation that emerges from IllustrisTNG ($M_{\rm BH} \propto M_{\rm halo}^{0.76}$) is much flatter than the one predicted by baryon lifting models. 

This paper therefore looked more deeply into the relationship between black hole mass and baryon lifting in IllustrisTNG, focusing on the TNG100 simulation, to determine the reason for the discrepancy. Previous work has already shown that quenching of star formation in IllustrisTNG is closely related to baryon lifting \citep{Davies_2020MNRAS.491.4462D,Terrazas2020MNRAS.493.1888T}, as reflected in a reduction of the halo gas fraction (Figure \ref{fig:M500-fgas-sSFR}). Throughout the redshift range $0 \leq z \leq 3$, the transition from active star formation to quiescence occurs when the cumulative kinetic energy input from black hole feedback becomes comparable to the halo's baryonic binding energy (Figure \ref{fig:M500-Ekin-sSFR}). Those findings imply that black hole mass growth during the period of star-formation quenching in IllustrisTNG is linked to baryon lifting in a manner consistent with the observed $M_{\rm BH}$--$T_{\rm CGM}$ relation. 

However, early black hole mass growth associated with thermal mode feedback in IllustrisTNG vastly exceeds later mass growth coinciding with baryon lifting. During a halo's early period of thermal mode feedback, the mass of its central black hole grows to exceed $\sim 10^8 \, M_\odot$ before much lifting occurs. That mass threshold is built into the switch that determines whether black hole feedback is in thermal mode or kinetic mode (see equation \ref{eq:BHmode}). The switch starts to favor kinetic feedback over thermal feedback as a halo's mass approaches $\sim 10^{12} \, M_\odot$ (Figure \ref{fig:M500-Ekin-BHmode}), where the observed ratio of $M_*$ to $M_{\rm halo}$ peaks. The onset of kinetic feedback therefore lifts the galaxy's atmosphere when the ratio of black hole mass to halo mass reaches $\sim 10^{-4}$ (Figures \ref{fig:M500-Ekin-BHmode}, \ref{fig:MBH-Ekin-BHmode}, and \ref{fig:MBH-Ekin-fgas}).  

As baryon lifting happens, the efficiency parameter for kinetic feedback ($\epsilon_{\rm kin}$) determines the associated amount of black hole mass growth. Its fiducial value in IllustrisTNG is $\epsilon_{\rm kin} = 0.2$, meaning that the black hole mass ``price" required to lift the galaxy's atmosphere and quench star formation is only $\sim 3 \times 10^5 \, M_\odot$ in a $10^{12} \, M_\odot$ halo (see equation \ref{eq:DeltaMBH}). That amount of mass growth is negligible compared to the black hole mass growth needed to switch on kinetic feedback. Subsequent black hole mass growth via mergers in IllustrisTNG therefore largely preserves the $M_{\rm BH} / M_{\rm halo}$ ratio that is in place at the time the kinetic mode comes to dominate. 

Reduction of the $\epsilon_{\rm kin}$ parameter in IllustrisTNG would increase the black hole mass price paid for baryon lifting in proportion to $\epsilon_{\rm kin}^{-1}$. The observed $M_{\rm BH}$--$T_{\rm CGM}$ relation implies a lower limit of $\epsilon_{\rm kin} \gtrsim 0.005$ \citep{DonahueVoit2022PhR...973....1D}. However, the parameter value employed in simulations may need to be greater because of inefficiencies in coupling between kinetic feedback and baryon lifting. For example, \citet{BoothSchaye_2010MNRAS.405L...1B} showed that setting the equivalent feedback parameter in their simulations to $\epsilon_{\rm fb} = 0.015$ maximized agreement with the $M_{\rm BH}$--$M_{\rm halo}$ relations inferred from the observations available at that time.

Initiating baryon lifting at a lower black hole mass in IllustrisTNG also seems necessary to improve agreement with observations. For example, the typical black hole mass in a star-forming galaxy with $M_* \sim 10^{10.5-11} \, M_\odot$ is observed to be $M_{\rm BH} \sim 10^7 \, M_\odot$, whereas $M_{\rm BH} \sim 10^8 \, M_\odot$ is typical for similar galaxies in IllustrisTNG \citep{Terrazas2020MNRAS.493.1888T}. Given the anticorrelation observed between $M_{\rm BH}$ and sSFR at fixed stellar mass \citep{Terrazas_2016ApJ...830L..12T,Terrazas_2017ApJ...844..170T}, it would appear that the majority of black hole mass growth in star-forming galaxies happens \textit{during} the quenching process, not prior to it \citep[e.g.,][]{ChenFaber_2020ApJ...897..102C}.

Exactly how to adjust the black hole feedback algorithm in IllustrisTNG remains an open question. The current algorithm initiates baryon lifting and star-formation quenching when $M_{\rm halo} \sim 10^{12} \, M_\odot$ and $M_{\rm *} \sim 10^{10.5} \, M_\odot$. In IllustrisTNG, the corresponding central black hole mass is $\sim 10^8 \, M_\odot$ near $z=0$ and $\sim 10^{8.5} \, M_\odot$ near $z=3$. Simply reducing the pivot mass in equation (\ref{eq:BHmode}) by an order of magnitude might bring about better agreement with observational constraints on the $M_{\rm BH}$--$M_{\rm halo}$ relation but would also substantially reduce the values of $M_*$ and $M_{\rm halo}$ at which quenching occurs. A different solution is therefore needed, one that limits black hole masses in star-forming galaxies to $\sim 10^7 \, M_\odot$ prior to quenching of star formation and allows them to rise to $\gtrsim 10^8 \, M_\odot$ as black hole feedback lifts the surrounding galactic atmosphere and shuts down star formation.

\begin{acknowledgements}
This paper is dedicated to the memory of Richard Bower, whose pioneering work on black-hole feedback and galactic atmospheres inspired many of the ideas we have presented. GMV is supported in part by grant AAG-2106575 from the NSF. BDO's contribution was supported by Chandra Grant TM2-23004X. The authors are grateful to Dylan Nelson, Annalisa Pillepich, and Nhut Truong, whose comments significantly improved the paper.
\end{acknowledgements}






\bibliography{precipitation}{}
\bibliographystyle{aasjournal}

\end{document}